\documentclass[aip,rsi,reprint,nofootinbib,longbibliography]{revtex4-2}
\usepackage{graphicx}

\usepackage{etoolbox}

\makeatletter
\appto\abstract{%
  \let\latexlist\list \rightskip=\leftskip
  \def\list{\edef\keeprightskip{\the\rightskip}\latexlist}%
  \patchcmd\latexlist{\ignorespaces}{\rightskip\keeprightskip\ignorespaces}{}{}%
}
\makeatother

\usepackage{physics}
\usepackage{listings}
\usepackage{amsmath}
\usepackage{xcolor}
\usepackage{dsfont}

\newcommand{\pyrpl}{PyRPL }

\begin{document}

\graphicspath{{graphics/}}

\title{FPGA-based feedback control of quantum optics experiments with the open source software package PyRPL}

\author{Leonhard Neuhaus}
\author{Micha\"{e}l Croquette}
\author{R\'emi Metzdorff}
\author{Sheon Chua}
\author{Pierre-Edouard Jacquet }
\author{Alexandre Journeaux}
\author{Antoine Heidmann}
\author{Tristan Briant}
\author{Thibaut Jacqmin}
\author{Pierre-Fran\c{c}ois Cohadon}
\author{Samuel Del\'eglise}
\affiliation{Laboratoire Kastler Brossel, Sorbonne Universit\'e, ENS - Universit\'e PSL, Coll\`ege de France,~CNRS, 4 place Jussieu, F-75005 Paris, France}

\date{\today}

\begin{abstract}
We present PyRPL, an open source software package that allows the implementation of automatic digital feedback controllers for quantum optics experiments on commercially available, affordable FPGA boards. Our software implements the digital generation of various types of error signals, from an analog input through the application of loop filters of high complexity and real-time gain adjustment for multiple analog output signals, including different algorithms for resonance search, lock acquisition sequences and in-loop gain optimization. Furthermore, all necessary diagnostic instruments such as an oscilloscope, a network analyzer and a spectrum analyzer are integrated into our software. Apart from providing a quickly scalable, automatic feedback controller, the lock performance that can be achieved by using PyRPL with imperfect equipment such as piezoelectric transducers and noisy amplifiers is better than the one achievable with standard analog controllers due to the higher complexity of implementable filters and possibilities of nonlinear operations in the FPGA. This drastically reduces the cost of added complexity when introducing additional feedback loops to an experiment. The open-source character also distinguishes PyRPL from commercial solutions, as it allows users to customize functionalities at various levels, ranging from the easy integration of PyRPL-based feedback controllers into existing setups to the modification of the FPGA functionality. A community of developers provides fast and efficient implementation and testing of software modifications. 
\end{abstract}
\maketitle

\section{Introduction}
Feedback control is ubiquitous in scientific experiments to minimize the effects of fluctuations and environment conditions. The theory behind feedback control, from the definition of ``plant'' and ``controller'', to transfer functions, phase delays and stability conditions... is well known\cite{Bechhoefer2005,BechhoeferBook}. So are the challenges implied: a properly designed controller has to diminish the effect of fluctuations with a large gain at Fourier frequencies where delay is insignificant, while the gain remains low enough at frequencies where delay is significant to avoid instabilities, without the controller introducing excessive delay itself. These challenges create a trade-off between the benefits that arise from introducing feedback into an experiment (stability, reproducibility) to the increase in costs and effort to achieve those controls.

The implementation of feedback controllers has long been performed with analog electronics and circuitry. However, cutting-edge experiments can be dynamic and complex, with changing requirements between measurements due to even minor differences in initial input parameters. This has motivated the movement towards digital feedback control \cite{Sparkes2011a, Huang2014, Seymour-Smith2010, Dietrich2009, Darsow2020}, using digital signal processing (DSP), software modules/modular structures, and re-programmable electronics. 
This allows flexibility and customization, as well as lowering costs and multiplying control loop circuits through duplication of software instead of hardware electronics. 
These developments are enabled by the growing availability of Micro-Controller Units (MCUs) and Field Programmable Gate Arrays (FPGAs). 

\begin{figure*}[t]
\centering
\includegraphics[width=17cm]{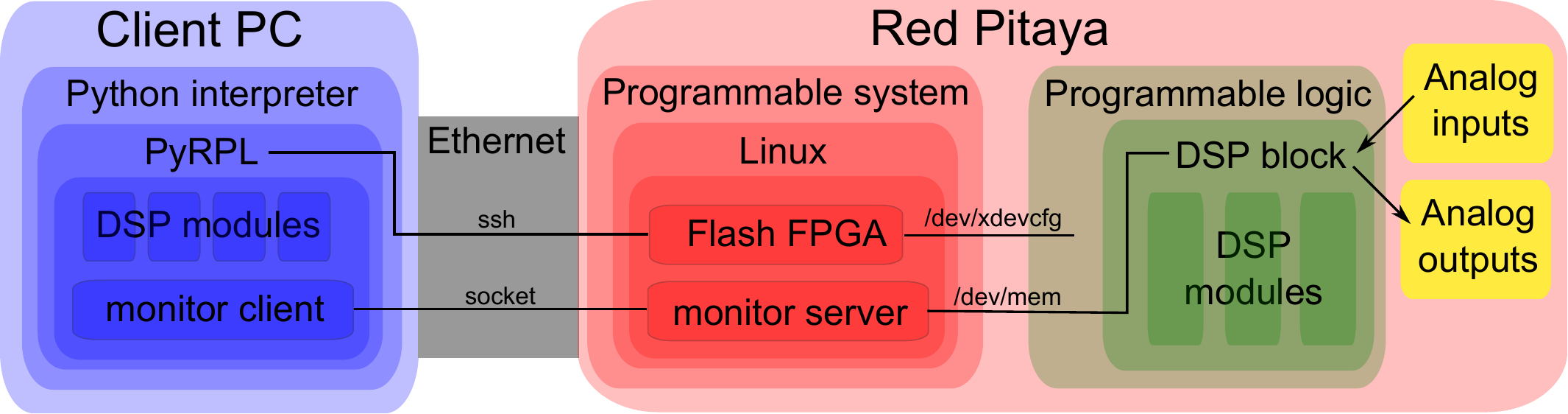}
\caption{\label{RPoverview} \small {\bf Overview of the PyRPL architecture}. It depicts the system components using different colors: blue blocks represent software written in Python, red blocks represent software in C running on the Red Pitaya's CPU, and green blocks represent Verilog code running on the FPGA chip. 
The system is controlled by a "Client PC" that incorporates a high-level programming layer, which may include an optional Graphical User Interface (GUI). 
During program startup, the Client PC establishes an SSH session with the Red Pitaya's operating system. It then proceeds to flash the FPGA board with the most recent version of the FPGA bin file, which contains the hardware description of the Digital Signal Processing (DSP) modules. 
Simultaneously, the Client PC initiates a "Monitor Server" on the Red Pitaya's programmable system. This server operates continuously, listening to the Client PC and ensuring synchronization between the DSP registers and attributes of Python objects that mirror the FPGA structure.}
\end{figure*}

Owing to their sequential logic, MCUs can be programmed and reconfigured easily in common programming language, such as C, making them a good choice when bandwidth requirements are not too tight. For instance, MCUs have been successfully used to stabilize laser-cavity systems with various locking algorithms \cite{Dietrich2009, Seymour-Smith2010, Huang2014}. On the other hand, FPGAs are advantageous when high locking bandwidth is required. They have been used in quantum optics experiments to implement complex control loops based on cascaded integrator comb filters \cite{Sparkes2011a}, Finite Impulse Response (FIR) \cite{Ryou2017} or Infinite Impulse Response (IIR) \cite{Okada2020} filters, or phase-lock loops \cite{Tourigny2019}. Furthermore, fast digital modulation and demodulation can be used to perform direct Pound-Drever-Hall signal generation and self-diagnose the loop with a digital network analyzer \cite{Sparkes2011a}.  
In contrast with MCUs, FPGAs demand a more specialized hardware description language and a lengthier, more intricate compilation process because of their parallel architecture. The technical subtleties involved in programming digital servo-loops with FPGAs \cite{Yu2018} is beyond the reach of many small research groups, who often struggle to maintain the necessary level of expertise over time. 
One way to leverage the power of FPGAs without the challenges of managing the hardware description language is to design a versatile system that can be reconfigured in real-time to suit diverse requirements, without recompiling the FPGA code. 
Furthermore, to maximize flexibility and customizability, and keep costs low, a digital controller should be built with mostly an open-source architecture. One such platforms has recently been developed with atomic spectroscopy applications in mind \cite{Wiegand2022}.

In this article, we present the \underline{Py}thon \underline{R}ed \underline{P}itaya \underline{L}ockbox open-source software package, or \textit{PyRPL}, based on the widely-available Red Pitaya field-programmable gate array (FPGA) platform, for digital feedback control. \pyrpl has been available in various evolving forms since 2016 and has already found its way in physics experiments in fields such as optomechanics \cite{rossi2018measurement,ruelle2022tunable,karg2020light},
quantum optics \cite{mazelanik2022optical,vaidya2020broadband,guo2020distributed,larsen2019fiber}
, atomic physics\cite{higgins2021micromotion,bhatt2022stochastic} and frequency combs\cite{barot2021mode}.

The idea behind the architecture of the program is to allow highly-flexible reconfiguration between DSP module connections, without the need for FPGA recompilation. This enables the user to switch instantly between different configurations, and to efficiently diagnose the control loop characteristics by connecting to acquisition modules (digital oscilloscope, spectrum analyzer, network analyzer - all available as parts of PyRPL). A Python program and Graphical User Interface (GUI), running on a separate computer, completes the Lockbox set to allow the monitoring of various signals and control of parameters and FPGA registers (via a TCP-IP connection) in real time, for full flexibility. This paper is divided into three sections. The first section presents an overview of the backbone architecture of PyRPL: the board, the client-computer communication and DSP multiplexer. The second section presents the various DSP modules within PyRPL. The last section presents example uses that call upon and arrange the various modules: Pound-Drever-Hall error signal generation, auto-feedback control sequences, and complex loops filters. 
The first section is mostly here for completeness and reference for the advanced reader. A reader only interested in a quick set-up of a given experimental platform is welcome to directly jump to the two latter sections. \\

\section{Python Red Pitaya Lockbox (PyRPL) Architecture}
A simplified schematic of the PyRPL architecture, with its key components is shown in Fig. \ref{RPoverview}. The Red Pitaya \cite{redpitaya} is an affordable FPGA board, with fast analog inputs and outputs, that has recently gained wide recognition in the scientific and electronics communities. PyRPL encompasses a user-interface and high-level functionality written in the Python language, client-computer communication protocol written in C language, and a background custom FPGA design in Verilog hardware description language for DSP multiplexing. The client computer hosts the first component for user flexibility and readout, while the Red Pitaya board hosts the latter two components for control loop implementation.

\subsection{Red Pitaya FPGA Board}
The Red Pitaya board combines a Xilinx Zynq-7010 System-on-chip (SoC) \cite{zynq} with two-channel analog-to-digital (ADC) and digital-to-analog (DAC) converters with 14 bit resolution and 125 MHz sampling rate. The Zynq-7010 combines a FPGA allowing programmable logic (PL) with a dual-core processing system (PS), composed of an ARM Cortex-A9 processor, and a high-bandwidth connection to the FPGA. The PS allows to run a Linux operating system (OS) on the board and thereby to use, interface, and program it like a standard personal computer. The PL is defined by a hardware description language, in this case Verilog, whose compiler generates a bitfile for the FPGA that can be loaded via the PS. The manufacturer of the Red Pitaya provides a sample Verilog source code that includes an example interface to the ADC and DAC, and an implementation of a bus between PS and PL that allows a Linux user connected to the board through an ethernet connection to read and write registers of the FPGA.

The PS clock rate is 1 GHz and the FPGA clock rate is 125 MHz. The FPGA clock is derived from an on-board quartz oscillator by default, but can be synchronized to an external clock. The DAC and ADC are clocked synchronous with the FPGA. However, feedback control loops are limited in bandwidth range to below 2 MHz, due to significant delay caused mainly by the employed ADC and DAC converter architecture.

The resolution of the ADC and DAC is 14 bits and both the differential nonlinearity and transition noise of both components are of the order of the least significant bit. However, the implementation of these components on the Red Pitaya board results in slightly worse performance. In agreement with previous work\cite{Seidler2015}, we have measured effective resolutions of 12 bits for the ADC and 11 bits for the DAC. 

\subsection{Client-computer communication}
The start up and connection to the Red Pitaya (with the starting of a PyRPL instance) begins with the link to the PS through a secure shell (SSH) connection. The client then uploads a bitfile which contains the compiled PL design and loads this file into the FPGA through a Linux command. Next, the client uploads and executes the C program "monitor\_server". The server program allows the client to connect to it through the TCP/IP protocol and listens for commands from the client. The possible commands are 'read' and 'write', followed by an address and the number of 32-bit packets of data to read or write from or to the FPGA, starting from the given address. If the address lies within the address space that the OS of the RedPitaya has assigned to the FPGA bus, the read or write request becomes available to the PL on the various bus signals. 

The requested command is interpreted by the logic implemented in the Verilog code, which essentially maps a number of the registers of each DSP module to specific addresses within an address subspace reserved for the module. By knowing the names and addresses of all registers of a DSP module, a Python object corresponding to the DSP module can be created on the client computer, such that the properties of the object correspond to the FPGA registers of the DSP module. When these properties are read or written, the Python value is automatically synchronized with the content of the corresponding registers through the communication chain. The end result is that a Python programmer can assign or read the FPGA registers in the same way as local Python variables. This provides rapid flexibility of the controller.

Communication speed has been a major concern in the development of PyRPL. Without excessive code optimization, we achieve typical read and write delays for a single 32-bit register of the order of 300 $\mu$s. When a block of registers with adjacent addresses are read or written, for example the memory storing an oscilloscope trace, the read/write delay scales sublinearly. For reading or writing $2^{14}$ 32-bit values, a time of the order of 7-10 ms is required. The delay is thus limited by the speed of the ethernet connection, and could be further improved by, for example, compressing the data sent over the network.

\subsection{DSP multiplexer}

\begin{figure}[tb]
\centering
\includegraphics[width=8.5cm]{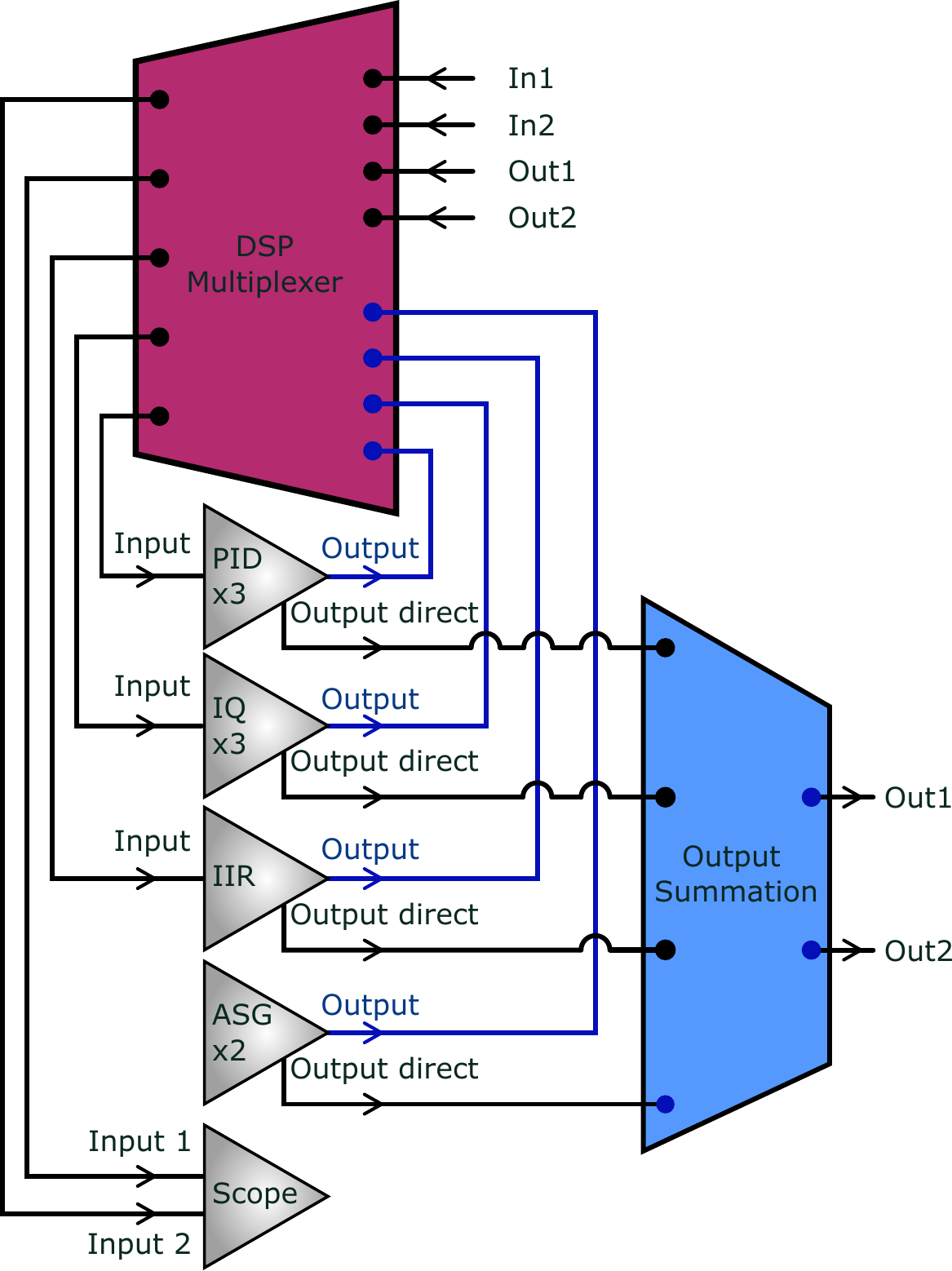}
\caption{\label{DSPblockfigure} \small 
{\bf Signal routing architecture of PyRPL}. The DSP modules are represented as triangles.
Each module, with the exception of the scope, has an output signal. 
A DSP multiplexer which functions as a signal router for the 14-bit signals traversing among the various DSP modules and the analog inputs (In1, In2) and outputs (Out1, Out2). 
Furthermore, each module (excluding the scope) boasts an additional "output\_direct" that can be directly routed to one of the analog outputs - "Out1" or "Out2" or both. All signals directed to the same analog outputs undergo summation before being dispatched to the Red Pitaya board's Digital-to-Analog Converter (DAC).}
\end{figure}

The DSP multiplexer serves as a router between the analog signals and the various DSP modules of PyRPL. Its Verilog implementation is shown in Fig. \ref{DSPblockfigure}. The module definition fits into a few lines of code, but is nevertheless very resource-intensive. The multiplexer can accommodate up to 16 DSP modules that each provide  a port for a 14-bit input signal ("input\_signal"), a 14-bit output signal ("output\_signal"), and an additional 14-bit signal called "output\_direct" which can be routed directly towards the analog outputs. The value of one register for each module called "input\_select" determines which other module's output signal is connected to the former module's input port. Another 2-bit register "output\_select" per DSP module is used to select whether the module's signal "output\_direct" is sent to the analog output 1, output 2, to both outputs or to none. As shown at the bottom of Fig. \ref{DSPblockfigure}, the sum of the selected "signal\_direct" of all DSP modules is computed for each of the two analog outputs and then sent to the DACs. The analog input signals from the ADCs and the output signals that are sent to the DACs are also represented in the form of DSP modules which only provide an "output\_signal" port in the DSP multiplexer in order to keep the interface simple. Some DSP modules have by definition only inputs or only outputs, and for others, the signals "output\_direct" and "output\_signal" are identical. 

By combining a large number of DSP modules in series, or by rapidly ($<1$ ms) modifying the connections between modules in real time, the DSP multiplexer allows the implementation of highly complex functionalities. Furthermore, it allows analysis modules (oscilloscope and analyzers) to be connected to any DSP module without the need of signal conversion to the analog domain, which has proven a highly practical tool for debugging, setup and in keeping analog cross-talk low. 

\section{Basic operation of the FPGA Modules}\label{DSPmodules}

As explained in the section above, the core of PyRPL is a set of FPGA modules that perform several DSP tasks, and that can be interconnected in various configurations by a DSP multiplexer without changing the FPGA bitfile. The various FPGA registers controlling the behavior of each FPGA module can be inspected and controlled transparently via a Python object mirroring each DSP module on the client computer. The state of the various modules (i.e. the value of the corresponding registers) are stored in a yml configuration file that is kept in-sync with the FPGA registers at all times. The state of each module can also be controlled via a GUI, and archived at any point for future usage.

The FPGA architecture is fully parallel and the modules are thus operating simultaneously on the digital signals sampled at the FPGA clock rate of 125 MHz. The default configuration of \pyrpl instantiates several copies of the following modules:
\begin{itemize}
\setlength{\itemsep}{0pt}
    \item dual channel cscilloscope 
    \item Arbitrary Signal Generator (ASG, $\times$2)
    \item Proportional Integrator Differential filter (PID, $\times$3)
    \item IQ modulator/demodulator ($\times$3)
    \item Infinite Impulse Response filter (IIR)\end{itemize}
This section discusses the principle of the various FPGA modules with emphasis on the underlying theory of DSP. Furthermore, we provide basic usage examples, together with full configuration instructions using the Python API \cite{notebook}.

\subsection{Data acquisition modules}
The possibility to monitor digital signals originating either from the input analog-to-digital converters of the Red Pitaya board or from any of the internal DSP modules is one of the key features of PyRPL, enabling fast debugging and prototyping of DSP setups. While reading or writing an FPGA register can be done in a synchronous manner---i.e the program execution on the computer is halted until the value has been successfully updated or read from the board---launching a data acquisition task, waiting for the data to be ready, and then displaying it on screen, cannot be done by a blocking program, otherwise, the GUI or other critical program components would be blocked for a long, and potentially unpredictable time. 

In PyRPL, we have decided to use the Qt event-loop framework to deal with asynchronous tasks. Compared to multithreading, an event-based approach avoids potential problems with race conditions and simultaneous memory access by concurrent threads. Thanks to the external library quamash, this approach is also compatible with the asynchronous framework \emph{asyncio} introduced natively in the Python syntax since version 3.3, thus greatly improving the readability of asynchronous codes. Since \pyrpl has been primarily designed for quick prototyping and testing, we have made sure that an interactive IPython shell (such as Jupyter notebook) can be used in conjunction with the GUI without blocking the data acquisition tasks.

\subsubsection{Oscilloscope}
The FPGA implementation of the two-channel oscilloscope is based on the example provided by Red Pitaya. The module records 2 synchronized time-traces of any DSP signals with $2^{14}$ consecutive points. An integer decimation register of the form $\{2^n\}_{0\le n \le 16}$ controls the number of consecutive samples averaged in each point, thus changing the total trace duration between 131 $\mu$s and 8.59 s. We have extended the triggering functionality and added various time-stamps which allows the user to keep track of the exact time of different traces. Furthermore, for trace durations above 0.1 s, a ``rolling-mode'' is available: in this mode, data are acquired continuously in a circular buffer and displayed at the maximum refresh rate on screen. This mode of operation is typically used for optical alignments or for the monitoring of slow drifts over time.

\begin{figure}[tb]
\centering
\includegraphics[width=8.5cm]{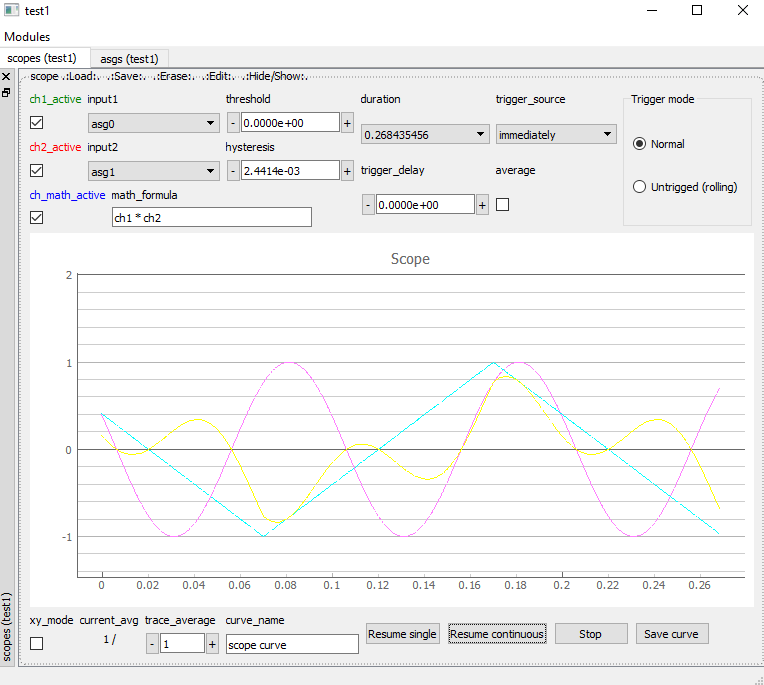}
\caption{\small {\bf Graphical user interface of the oscilloscope module}. The FPGA registers controlling the scope operation can be modified either from the controls located on the top panel or via an interactive python shell. In the latter case, the graphical displays are automatically synchronized with the updated value.}
\label{scopeGuis}
\end{figure}

Apart from mirroring the various registers necessary to control the operation of the FPGA oscilloscope (such as the trace duration, input channels, trigger source and thresholds...), the corresponding 
Python module exposes some high-level methods to acquire individual or averaged traces.
The fast communication interface presented above allows the display of both oscilloscope channels in real time on the client computer with a typical refresh rate exceeding 20 frames per seconds (see the GUI in Fig. \ref{scopeGuis}). 

\subsubsection{Spectrum analyzer}
By performing the FFT of individual oscilloscope traces, the spectrum analyzer can display in real-time the spectrum of the ADC inputs or internal DSP signals. The spectrum is estimated by implementing the Welsch method \cite{Welch67}: the user can pick a filter window among several options: 'blackman', 'flattop', 'boxcar', 'hamming', or 'gaussian'. The span, given by the scope sampling rate $1/T_\mathrm{scope}$, is selected among a list of predefined values. Equivalently, the residual bandwidth $\mathrm{RBW}$ can be chosen from a predefined list, via the relation
\begin{equation}
    \mathrm{RBW} = \mathrm{ENBW}/T_\mathrm{scope} ,
\end{equation}
where the equivalent noise bandwidth (ENBW) depends on the filter function samples $\{f_n\}_{0\le n < N}$ via $\mathrm{ENBW} = \left(\sum_n^{N}{f_n^2}\right)/\left(\sum_n^{N}{f_n}\right)^2$. The spectrum analyzer has two modes of operation: in ``baseband'' mode, the FFT operations are performed on the 2 real channels of the oscilloscope, allowing us to extract the individual real spectra and the complex cross-spectrum of the 2 input signals. In ``IQ-mode'', a single input signal is demodulated around a central frequency $f_0$ by an IQ-module and the 2 slowly varying quadratures are merged to form a complex time series $z_n = I_n + j Q_n$. By performing the FFT of this complex time series, we extract the spectrum of the input signal in a narrow window around the central frequency $\nu_0$.

\subsection{Arbitrary signal generator module}
The arbitrary signal generator module (ASG) module is an adapted version of the two-channel ASG provided by the Red Pitaya manufacturer. Our Python interface allows to slightly reduce the necessary FPGA resources for the ASG in order to reserve more FPGA space for other modules. Waveforms defined by $2^{14}$ values are easily loaded through the Python interface. The ASG supports frequencies from 0.1 Hz to 62.5 MHz and various burst and pulse modes. We have added an extra triggering functionality to the ASG in order to allow arbitrary delays for its turn-on or turn-off with respect to the arrival time of an external trigger signal. 
Furthermore, we have implemented a pseudo-random noise generator (PRNG) based on a Lehmer PRNG \cite{Payne1969}. While the current generator has a period of the order of 2 s, true random number generation is possible with FPGAs \cite{baetoniu2008method} and might be implemented in the future.

\subsection{Proportional-Integrator-Differential module}
The Proportional-Integrator-Differential (PID) module is an extended version of a Red Pitaya example. The output signal $s_\mathrm{out}$ computed by a standard PID module with integral gain $g_\mathrm{i}$, proportional gain $g_\mathrm{p}$ and derivative gain $g_\mathrm{d}$ can be written as
\begin{equation}
s_\mathrm{out}(t) = g_\mathrm{i}\int_{-\infty}^{t}e(\tau)\mathrm{d}\tau + g_\mathrm{p}e(t) + g_\mathrm{d}\frac{\mathrm{d}}{\mathrm{d}t} e(t) \, ,
\end{equation}
where $e = s_\mathrm{in}(t) - s_0$ is the difference between the input signal $s_\mathrm{in}$ and the setpoint $s_0$. A very practical modification of the PID module is the possibility to modify the current value of the integral in the above equation through a register termed "ival" that is accessible in the Python interface. This way, the PID integrator can be re-set to arbitrary output voltages, or voltage ramps can be easily implemented as for-loops in Python that periodically change the value of the integral. 

In addition to this, we have added 4 first-order filters with selectable low-pass or high-pass cut-off frequencies in series to pre-filter the input signal of the PID module. An additional saturation stage allows to define arbitrary maximum and minimum voltages for the PID output signal. We note that, while the saturation stage is often practical to protect sensitive components connected to a Red Pitaya output from undesired voltages, it is always preferable to build an analog attenuator circuit instead of using digital saturation when possible, in order to benefit from a maximum dynamic range and a better analog noise performance. Similarly, it is always preferable to replace digital low-pass filters acting on the output signal of a Red Pitaya by analog ones in order to improve the averaging of the DAC output noise.

\subsection{IQ module}
\label{IQ_FPGA}
The linear response of a system subjected to a sinusoidal perturbation $V_\mathrm{exc}(t) = A_\mathrm{exc}\cos(\omega_0 t)$ is a signal at the same frequency: $V_\mathrm{meas}(t) = H(\omega_0) A_\mathrm{exc} \cos(\omega_0 t + \phi(\omega_0))$ that can be decomposed in two orthogonal components $V_\mathrm{meas}(t) = I(\omega_0) \cos(\omega_0 t) + Q(\omega_0) \sin(\omega_0 t)$. In practice, the $I$ (resp. $Q$) quadratures can be retrieved by applying a low-pass filter on the product $V_\mathrm{meas}(t) \cos(\omega_0 t)$ (resp. $V_\mathrm{meas}(t) \sin(\omega_0 t)$). This so-called IQ demodulation is the core operating principle of vector network analyzers and forms the basis of various sensitive measurement technics such as lock-in detection or the Pound-Drever-Hall error signal generation  widely used to stabilize a laser frequency to an optical cavity \cite{Drever1983}. While this vector demodulation is routinely achieved with analog electronic components, the IQ module of \pyrpl allows to perform this operation using the digital signal processing capabilities of the FPGA board provided the excitation frequency $\omega_0/2\pi$ doesn't exceed the Nyquist frequency of the analog-to-digital converters ($\omega_{N}/2\pi$ = 62.5 MHz). Generating these error signals numerically is advantageous as analog mixers usually suffer from imperfections such as DC-offsets that directly translate into the generated error signal. 
Furthermore, in phase locked loop applications, it can be advantageous to extract an error signal proportional to the phase of the slowly evolving quadrature signal. This phase can be estimated and unwrapped in real time thanks to a CORDIC phase detection algorithm implemented in the IQ module of PyRPL.

Finally, for an incoherent input signal $V_\mathrm{meas}(t) = \Re\left(\int_{0}^{\infty} V(\omega) e^{i \omega t} d\omega/2 \pi \right)$, the IQ demodulation procedure provides a frequency-shifted version of the signal:
\begin{equation}
I(t) + j Q(t)  \approx  \int_{0}^{\infty} V(\omega - \omega_0) e^{i\omega t} H_f(\omega)  d\omega/2 \pi,
\end{equation}
where the transfer function $H_f(\omega)$ of the low-pass filter is assumed to have a cutoff frequency $\omega_f \ll \omega_0$. A bandpass filtered version of the input signal can then be generated by applying the reverse IQ modulation:
\begin{align*}
V_{out}(t) &= I(t) \cos(\omega_0 t) + Q(t) \sin(\omega_0 t) \\
&= \int_{0}^{\infty} V(\omega)  H_f(\omega - \omega_0) e^{i \omega t}  d\omega/2 \pi.
\end{align*}

\label{IQmodule}
\begin{figure}[tb]
\centering
\includegraphics[width=8.5cm]{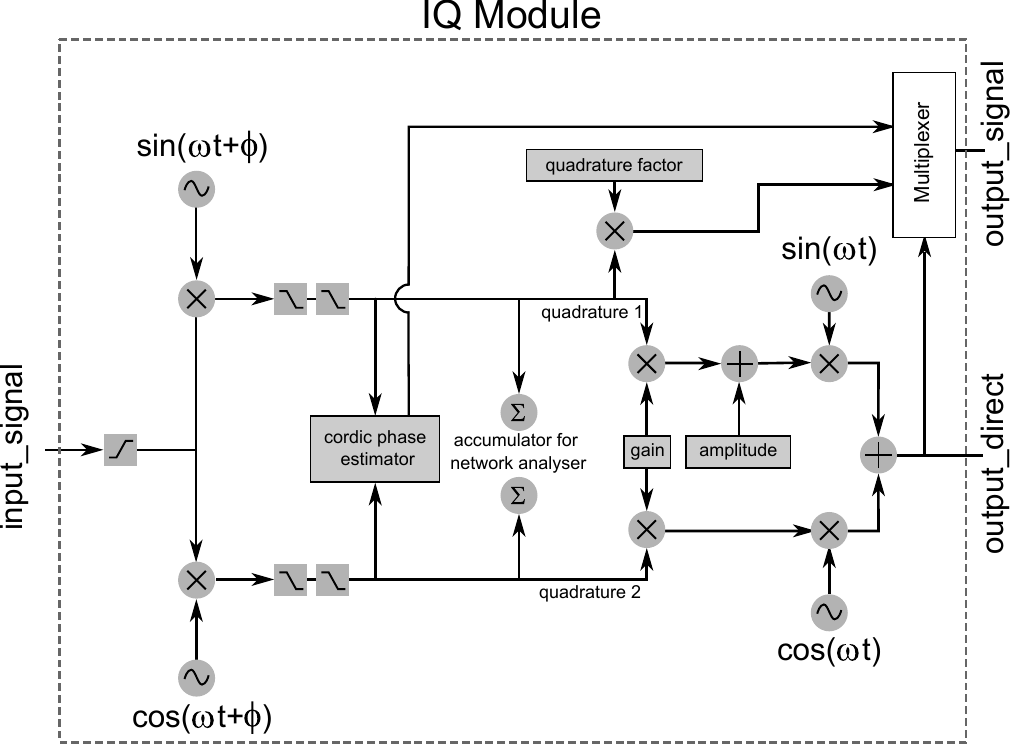}
\caption{\small {\bf IQ module layout}. The input signal is first high-pass filtered with a cutoff frequency defined by the register "ac\_bandwidth". The signal is then demodulated at a frequency defined by the register "frequency", with a phase defined by the "phase" register. The slowly varying quadratures are then obtained by applying second order low-pass filters on the multiplier's outputs (cutoff controlled by the register "bandwidth"). The quadrature signals can then be used for various applications: they are sent to a CORDIC phase estimator to extract a phase-lock error signal, accumulated in high-precision registers for the network analyzer, or directly used as the module's "output\_signal" depending on the value of the register "output\_signal". 
}
\label{IQmodulepic}
\end{figure}

The design of the IQ module is schematized in Fig. \ref{IQmodulepic}. 
The four phase-shifted sine functions required for each IQ-module are generated from a look-up table (LUT) of $2^{11}$ 17-bit values stored in read-only memory (ROM) of the FPGA. For minimum resource requirements per IQ module, only a quarter period of the sine function is stored in the LUT. The phase of the sines is given by the most significant bit of a 32-bit register that is incremented each clock cycle by the value of the register that defines the frequency.

Even though the wiring is fixed by design in the FPGA code, the module can be instantly reconfigured for various applications, by setting some of the registers ``gain'', ``amplitude'', and ``quadrature\_factor'' to zero, effectively disabling the corresponding signal paths. An output multiplexer ''output\_signal'' is also present to afford more flexibility. In the following, we provide examples and practical configuration instructions for various modes of operation: as a vector network analyzer, as a tunable bandpass filter and for Pound-Drever-Hall error signal generation.

\begin{figure*}[t]
    \centering
    \includegraphics[width=\textwidth]{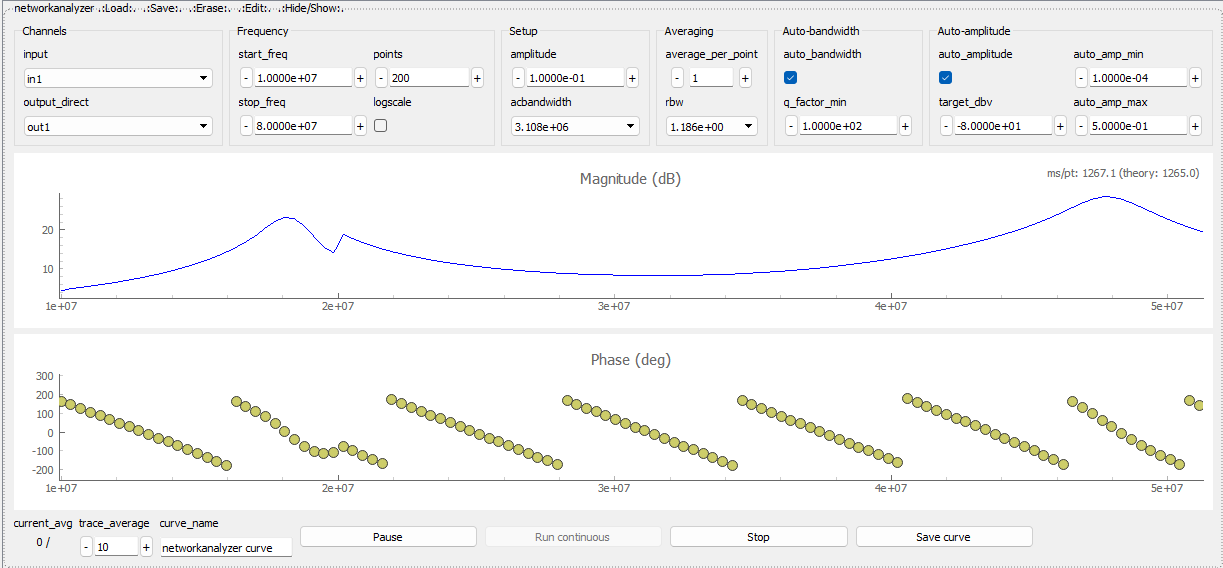}
    \caption{{\bf Graphical User Interface of the network analyzer}. The user can select the input and output of the network analyzer with the corresponding registers (within the "Channels" group). The frequency sweep is defined by the parameters "start\_freq", "stop\_freq" and "points" within the "Frequency" group. When the logscale checkbox is ticked, the frequency-bins are logarithmically distributed. The "amplitude" and "acbandwidth" control output voltage and high-pass filter cutoff at the input of the iq-module. The "rbw" and "average\_per\_point" registers determine the averaging time for each point. When "auto\_bandwidth" is ticked, the residual bandwidth is optimized at each frequency bin. "auto\_amplitude" is used to stabilize the input amplitude based on the value of the transfer function measured on previous points.}
    \label{fig:na}
\end{figure*}

\subsubsection{Network analyzer}
In order to use an IQ module as a network analyzer, the register "gain" is set to zero, effectively decoupling the demodulation and modulation stages of the module (see Fig. \ref{fig:na}). The "amplitude" register is set to a non-zero value governing the amplitude of the sinusoidal excitation at the "output\_direct" of the module. This excitation can be routed either to an analog device-under-test (such as an electro-optic modulator or a piezo-actuator) by setting "output\_direct" to "out1" or "out2", or to an internal DSP module, as demonstrated in \ref{sec:band_pass}.

After going through the device-under-test, the signal is connected to ``input\_signal'', and  optionally high-pass filtered at the beginning of the IQ module. The signal is then demodulated at the modulation frequency. One or two subsequent low-pass filters remove the demodulation component at twice the modulation frequency before being accumulated in two registers corresponding to the real and imaginary parts of the transfer-function. 
The sweep over a list of frequencies is performed by a for-loop on the python client.  Updating the frequency register of the FPGA module\footnote{In zero-span mode, the frequency register is updated with its current value, which also triggers the acquisition of a new network analyzer value.} triggers a counter of a pre-defined number of cycles specified by a register "sleep\_cycles". This is done to allow for the settling of the system under test before the acquisition of its transfer function begins. Next, during a number of cycles defined by a register "na\_cycles", the values of both demodulated quadratures are additively accumulated inside the two 62-bits accumulators shown in Fig. \ref{IQmodulepic}. Once the "na\_cycles" counter reaches zero, the python client reads the values of the two accumulators as 62-bit numbers and computes the averaged complex transfer function.

The configuration of the IQ module and the data-acquisition loop is handled by a pure software module that exposes high-level functionalities such as setting the start and stop frequencies of the scan, or the IF bandwidth of the measurement, as well as curve acquisition functions identical to those of the scope and spectrum analyzer. The GUI interface of the network analyzer is represented in Fig. \ref{fig:na}.

\begin{figure*} [t]
    \centering
    \includegraphics[width=\textwidth] {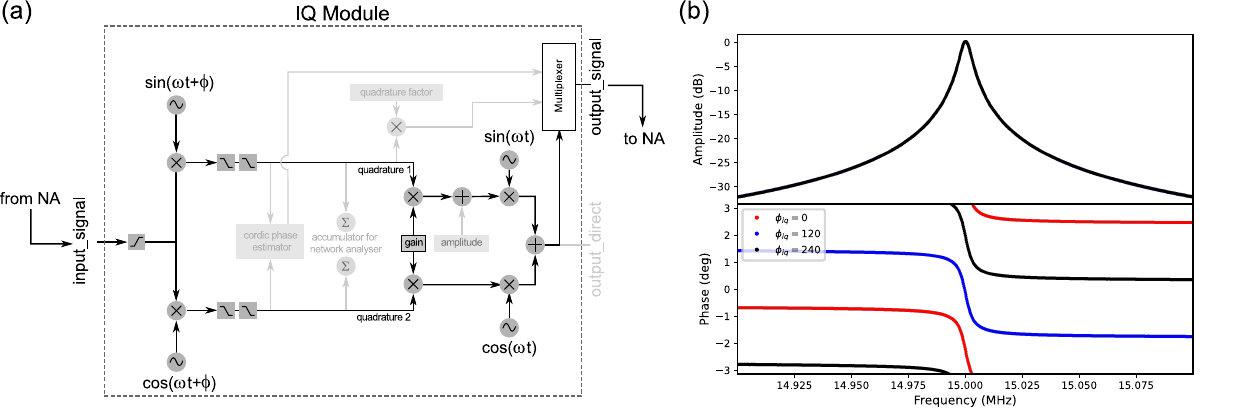}
    \caption{{\bf IQ module as bandpass filter}. (a) Schematic of the IQ module configured for bandpass filter implementation. Signal paths that are deactivated by setting the register values "quadrature\_factor" and "amplitude" to 0 are dimmed in the diagram.
    (b) Bode plot of the bandpass filter measured with the internal network analyzer of \pyrpl for 3 values of the phase register (0$^\circ$, 120$^\circ$, 240 $^\circ$), a demodulation frequency of 15 MHz, and a first order low-pass filter of bandwidth 2.3 kHz.}
    \label{fig:bandpass}
\end{figure*}

While the analog performance of the network analyzer does not compete with commercial network analyzers, a number of custom adjustments often allow better functionality.
For example, by iterating over a list of frequencies in reversed order, frequency sweeps in both frequency directions have been performed. This feature is commonly inaccessible in commercial network analyzers, and can be used for instance to probe hysteretic behaviours such as Duffing non-linearities in mechanical systems.

Another practical customization involves the stabilization of the magnitude of the input signal by adjusting the amplitude of the output tone, which was found to yield better measurement results in systems with response of high dynamic range and nonlinear response, such as high-$Q$ mechanical resonances detected with a Fabry-Perot cavity. Finally, the simultaneous availability of network analyzer and feedback control modules  inside a single DSP system makes \pyrpl a convenient tool to probe the in-loop transfer function of a closed-loop system, as demonstrated in section \ref{sec:in_loop_transfer}. This ability is particularly useful when working with optical Fabry-Perot cavities, where a stable lock is mandatory to measure any optical response. 

\subsubsection{High-Q band-pass filter}
\label{sec:band_pass}
In this example, an IQ module is employed to realize a band-pass filter. 
The transfer function of the implemented filter is measured in-situ by the integrated network analyzer of PyRPL, without transiting in the analog domain. 
Since this example doesn't require any external hardware, it can be seen as a basic tutorial for early users to gain hands-on experience with \pyrpl.

For the band-pass filter operation mode of the IQ module, the value of the "amplitude" register is set to zero and instead the "gain" register is set to the desired filter gain. An incoming signal is then demodulated with a phase-shifted $\sin$/$\cos$-pair, low-pass filtered, and modulated with the un-shifted $\sin$/$\cos$-pair (see Fig. \ref{fig:bandpass}a). The center frequency of the filter is thus determined by the frequency register of the IQ module, while the phase lag can be tuned by the relative phase-shift between the demodulation and modulation oscillators \cite{Verlot}. The filter bandwidth is defined by the cut-off frequencies of the low-pass filters acting on the demodulated quadratures. If only one low-pass filter is used, the shape of the band-pass filter is Lorentzian. 

By connecting the implemented filter to the integrated network analyzer of \pyrpl 
(internally employing another IQ module), it is possible to measure the filter's transfer function. 
Figure \ref{fig:bandpass}b shows the transfer function measured for a "gain" register of 1, a first-order filter of bandwidth 2.3 kHz and a set of three different phases $\phi = 0^\circ$, $120^\circ$ and $240^\circ$. 

By combining up to three (limited by the number of IQ modules) band-pass filters of the described type in parallel, complex transfer functions can be realized. The digital quadratures of narrow band-pass filters can benefit from significant reduction of the ADC noise through averaging and are therefore internally represented with 24 instead of 14 bits. This advantage is easily lost through the noise added by the DACs. For optimal performance, it is therefore crucial to adjust the filter gain such that the output signal nearly saturates the available voltage range, and to use analog attenuation afterwards, which attenuates both the signal and the DAC noise.  

\subsubsection{Pound-Drever-Hall error signal generation}

\begin{figure*}
    \centering
    \includegraphics[width=\textwidth] {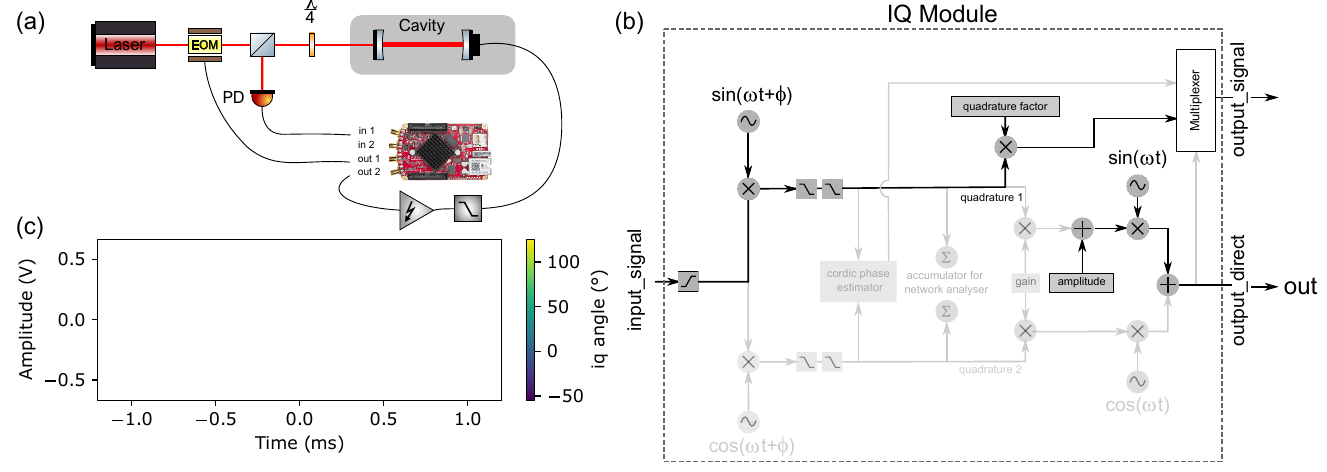}
    \caption{{\bf Pound-Drever-Hall error signal generation}. (a) Experimental setup: a laser is sent into an optical cavity. An electro-optic modulator (EOM) connected to the output "out1" of the Red Pitaya generates a phase modulation. When the laser is detuned from the cavity resonance, this creates an intensity modulation, that is detected by a fast photodiode (PD), digitized by the input "in1", and subsequenctly demodulated to extract a dispersive error signal.   
    (b) The phase modulation tone is generated by the "output\_direct" of the IQ module, with an amplitude of 1 V, and a frequency of 50 MHz. The PD signal is demodulated at the same frequency, and low-pass filtered with a second-order cutoff frequency of 3 MHz. (c) Error signal obtained by sweeping the cavity length for various IQ demodulation phases. The optimal demodulation phase, for which the slope of the error signal is maximum, is 35 $^\circ$.}
    \label{fig:pdh}
\end{figure*}

In order to use the IQ module for the generation of a Pound-Drever-Hall (PDH) error signal, the register ``gain'' is set to zero, the registers ``amplitude'' and ``quadrature\_factor'' to non-zero values and the output multiplexer ``output\_signal'' is  set to ``quadrature'' (see Fig. \ref{fig:pdh}b). With this setting, the signal ``output\_direct'' features a sinusoidal modulation that can be sent through one of the analog outputs to a modulator such as an EOM or a piezoelectric actuator. In the example depicted in Fig.
\ref{fig:pdh}, a phase modulation is imprinted on the laser via an EOM. 
The detected analog signal is digitized and connected to "input\_signal". The signal is optionally high-pass filtered at the beginning of the IQ module and then demodulated. One or two subsequent low-pass filters remove the demodulation component at twice the modulation frequency before an amplified version of one demodulated quadrature is available as "output\_signal" of the IQ module.

Fig. \ref{fig:pdh}c shows the generated PDH error signal, together with the reflection as the cavity length is swept across the resonance via a piezo-electric actuator. The figure also shows how the error signal amplitude is optimized by varying the demodulation phase.

\subsection{Infinite Impulse Response module}
\label{IIRsection}
A common problem in practical control applications is to compensate for the complicated transfer function of an imperfect actuator. A prominent example from optical cavity stabilization are the sharp resonant features arising from mechanical resonances of a piezoelectric actuator. While simple filters can be realized by combining several PID and IQ modules, the shape and complexity of the filters that can be constructed with a small number of modules is limited. On the contrary, digital infinite impulse response (IIR) filters \cite{} can be used to generate quite complex transfer functions with a relatively small number of operations per clock cycle. We  present here the implementation of an IIR filter in PyRPL.
A more detailed theoretical description of the IIR filter is given together with its practical implementation in Appendix \ref{AppendixIIR}.

A linear shift-invariant digital filter can be  characterized by its impulse response $h(n)$: \begin{equation}
y(n) = \sum_{i=0}^\infty h(i)x(n - i), \\
\label{convolutionsum}
\end{equation} 
where $y(n)$ is the output signal and $x(n)$ the input signal. All time-dependent signals are here referenced to the discrete sampling times $t=n T$, with $T$ the inverse sampling rate.
 A useful representation of the impulse response $h$ is given by its Z-transform:
\begin{equation}
H(z) = \sum_{n=0}^{\infty} h(n) z^{-n}.
\label{z-transform}
\end{equation}
For all practical filters, the summation in Eq. (\ref{convolutionsum}) is limited to a few cycles $i$ (typically 2 for this article), corresponding to a \emph {Finite Impulse Response} (FIR) filter.

A more general IIR filter, which corresponds to an infinite number of non-zero terms, can be mimicked with a few cycles and an additional dependence on the former values of the output:  
\begin{align}
y_j(n) =& b_0 x(n) + b_1 x(n-1) \nonumber\\
&- a_1 y_j(n-1) - a_2 y_j(n-2),
\end{align}
with the corresponding $H$ transform:
\begin{align}
H(z) &= D +\sum_{j=1}^{N'} \frac{b_{0j} + b_{1j} z^{-1}}{1 + a_{1j} z^{-1} + a_{2j} z^{-2}},
\label{biquads_decomposition}
\end{align}
where $D$ and all $a$ and $b$ coefficients are real. The overall filter therefore appears as a sum of simple filters with zeroes and poles.

\begin{figure*}[t!]
    \centering
    \includegraphics[width=\textwidth]{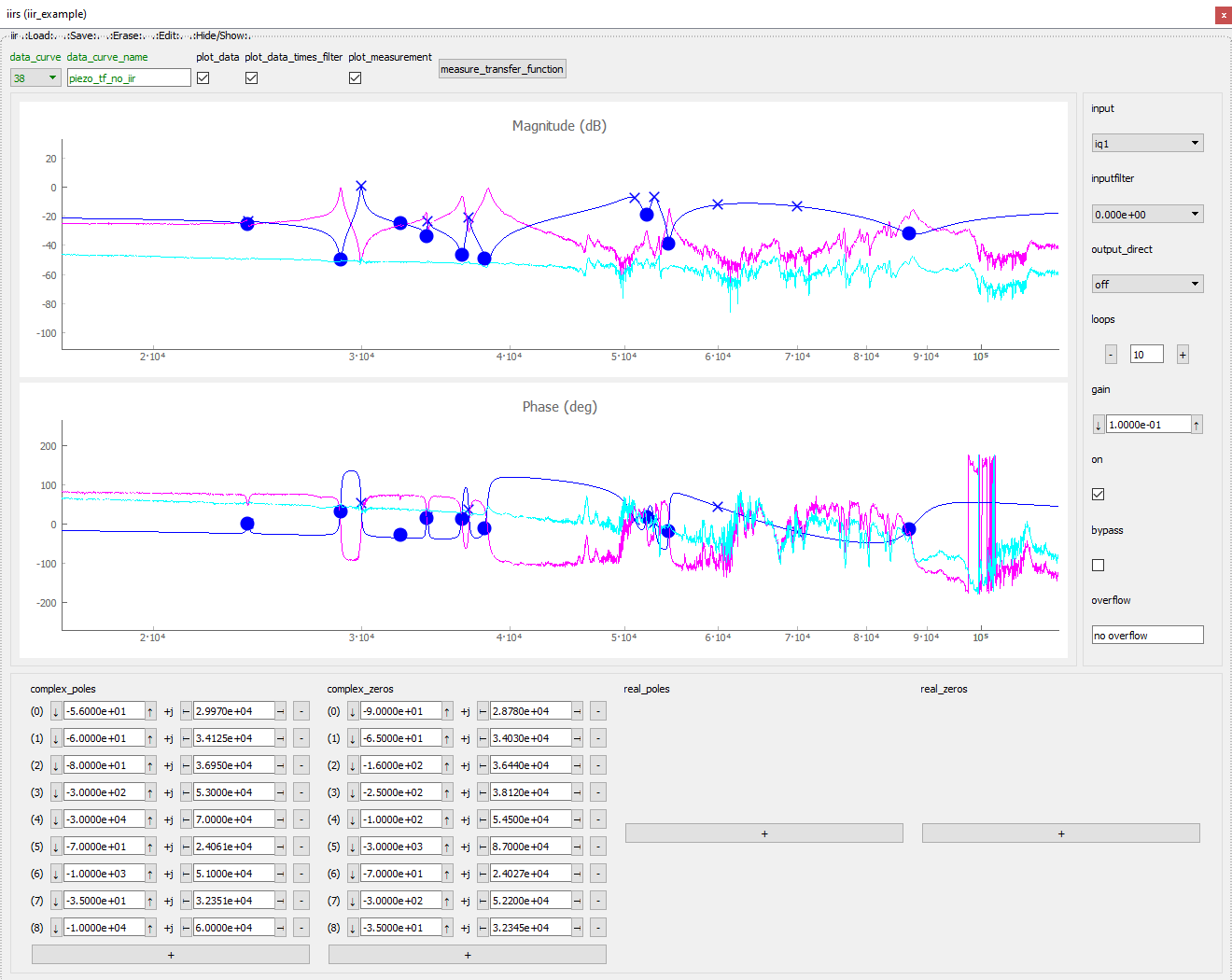}
    \caption{{\bf Graphical User Interface of the IIR module}. 
    To help with the design of the filter, the measured actuator response curve can be selected in the upper panel. The actuator response (pink), current filter response (blue), and their product (cyan) are represented in the Bode plot below. 
    A response as flat as possible in both amplitude and phase is usually achieved by adding zeros and poles in the lists at the bottom of the screen. 
    The rightmost panel allows to choose the input, optional filter, ``output\_direct'', the downsampling factor (loops), and overall gain of the filter.}
    \label{fig:iir_gui}
\end{figure*}

Fig. \ref{fig:iir_gui} shows the GUI of the IIR module. A graphical representation of the total transfer function (IIR filter + analog plant) helps the user to iteratively optimize the controller by adding poles and zeros to the IIR filter design. Typically, transfer functions with more than ten zeros and poles can be implemented with a negligible delay up to several hundred kilohertz. We provide a concrete example of loop optimization using the IIR filter in section \ref{sec:iir_experiment}.

\section{Applications of \pyrpl  in quantum optics experiments}

In this section, we provide several example applications, where the capabilities of \pyrpl are exploited to solve real-life laboratory problems. The applications, sorted from the simplest to the most complex, cover a wide range of experimental problems that are ubiquitous in laser physics and quantum optics laboratories:
\begin{itemize}
    \item Configuration of an automated lock-acquisition sequence for an optical Fabry-Perot cavity.
    \item milliradian precision stabilization of 2-independent lasers with a digital phase-lock-loop.
    \item Measurement of the transfer-function of a feedback-loop used to stabilize a high-finesse cavity up to 500 kHz.
    \item Active compensation of the acoustic resonances of a piezo-electric actuator using PyRPL's IIR filter.
\end{itemize}
In order to enhance the reproducibility of these results and enlarge the adoption of \pyrpl among the community, we provide, for each application, a Jupyter notebook demonstrating the step-by-step configuration of \pyrpl towards the desired goal  \cite{notebook}.

\begin{figure*}[t!]
    \centering
    \includegraphics[width=\textwidth]{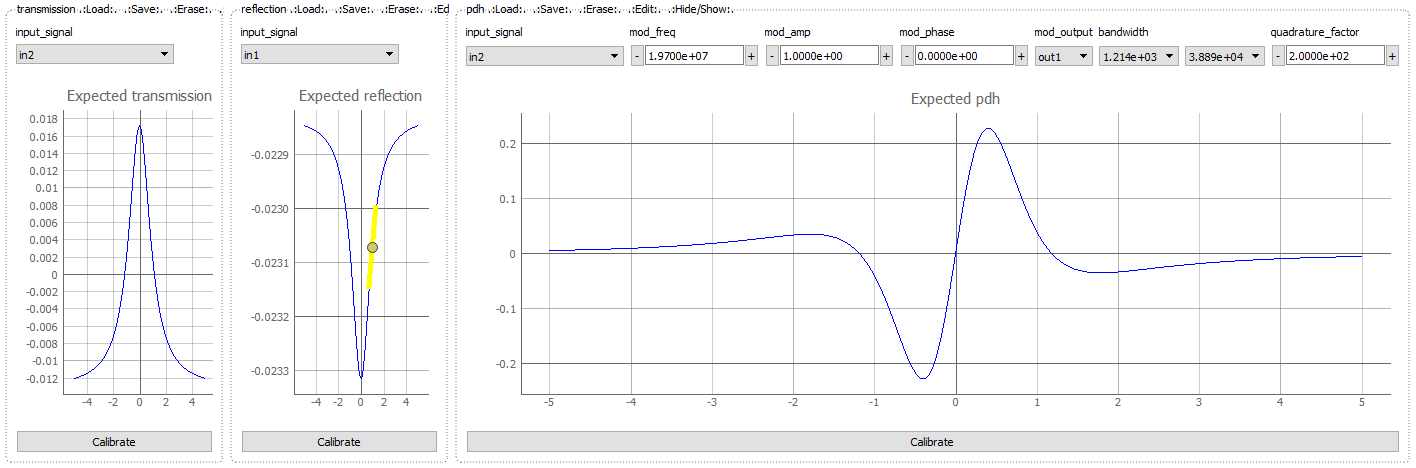}
    \caption{{\bf GUI panel describing the various inputs of the FabryPerot Lockbox}. In this example, the transmission signal is connected to the analog input ``in2'' of the Red Pitaya, the reflection signal is connected to ``in1'', and the PDH error signal is generated thanks to an EOM connected to ``out1'' (see Fig. \ref{fig:pdh}). The parameters for the modulation/demodulation, directly editable in the ``pdh'' input panel are passed transparently to the underlying IQ module. The button ``calibrate'' below each plot can be used to determine the offset and amplitude of the corresponding error signal experimentally.}
    \label{lockbox_inputs}
\end{figure*}

\subsection{Fabry-Perot lock acquisition}

With analog electronics systems, the lock acquisition sequence is often a tedious manual process, involving several steps, such as manually approaching the resonance, switching on the integrator gain, increasing the proportional gain... The situation is even more dramatic when several optical elements are placed in series: then a loss-of-lock of a single element requires a manual intervention to relock all subsequent elements in the chain. 
In contrast, the possibility to control and reconfigure the DSP modules remotely with \pyrpl enables to fully automate the lock acquisition sequence. 

In order to quickly configure and optimize lock-acquisition sequences, we have developed a purely software module (with no direct counterpart in the FPGA board): the Lockbox module. A specific lockbox type, containing a list of inputs and outputs, is defined for each sort of optical system by subclassing the Lockbox class. 
For instance, the FabryPerot Lockbox contains ``reflection'', ``transmission'', and ``pdh'' error signals, all displayed on Fig. \ref{lockbox_inputs}.

Since the expected signal shape $x_e(\Theta)$ is explicitly provided by the model, it is possible to stabilize the system at different setpoints $\Theta$ (in the example of the FabryPerot Lockbox, $\Theta$ is the cavity detuning). Furthermore, since optical elements are often subject to misalignments and loss of contrasts, we  typically allow for a vertical offset $y_0$ and scaling factor $g$ in the actual error signal $\langle x_e\rangle (\Theta)$. These parameters are experimentally determined  by recording the error signal while sweeping the actuator over the relevant range. This calibration procedure is fully automated such that it can be repeated whenever an alignment change is suspected\footnote{The overall gain of the loop is specified in dimensionless units, such that the controller gain can be adjusted to compensate for variations of the slope $\frac{\partial x_e(\Theta)}{\partial\Theta}|_{\Theta_0}$.}. 

\begin{figure}
    \centering
    \includegraphics[width=0.48\textwidth] {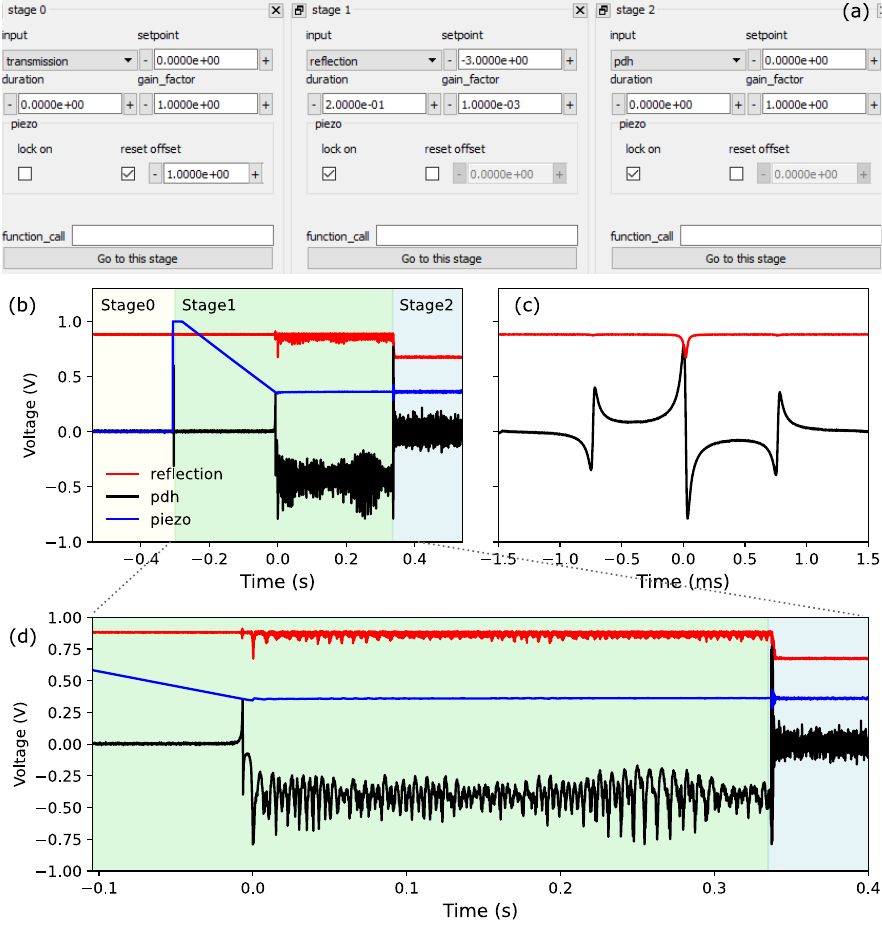}
    \caption{{\bf Complex lock acquisition sequence}. (a) As part of the Lockbox module, a locking sequence is defined as a succession of steps, which can use different locking schemes configurable via the GUI. 
    (b) Evolution of the reflection and PDH error signals, together with the piezoelectric actuation signal as a function of time, during the lock-acquisition sequence. Stage 1 consists in a "side-of-the-fringe" lock with a detuning of -3 cavity bandwiths, during which an integrator is drifting until the cavity resonance condition is reached. 
    At stage 2, the controller switches to PDH stabilization at the cavity resonance.
    (c) Reflection and PDH error signal shapes acquired during a sweep of the cavity length (with the same y-scale). 
     (d) Zoom on the central part of trace (b). }
    \label{fig:lock_sequence}
\end{figure}

We have found that a simple yet robust lock-acquisition could be achieved on most optical systems thanks to a generic sequential process: the lock-acquisition sequence is divided into successive stages, where each stage activates a specific feedback loop (see Fig. \ref{fig:lock_sequence}). 
For each stage in the sequence, the user specifies the error signal $x_e(\Theta)$ to stabilize, and the specific setpoint $\Theta_0$. 
Switching between various error signals proves to be useful since some error signals commonly feature multiple fixed points $\{\Theta_i\}_{0\le i< N}$ such that $\langle x_e \rangle (\Theta_i) = \langle x_e \rangle (\Theta_0)$. On the other hand, by choosing $\Theta_0$ on a per-stage basis, the user can avoid singular points where $\partial \langle x_e \rangle / \partial \Theta = 0$. 

This example demonstrates how the extra abstraction layer introduced by the Lockbox module can be used to configure a sequence to approach the resonance and acquire the lock reproducibly on a high-finesse optical cavity. 
Fig. \ref{fig:lock_sequence}a is a screenshot of the widget used to define the various stages in the lock-acquisition sequence. We record and display in Fig. \ref{fig:lock_sequence}b the evolution of the various error signals (reflection in red, pdh in black) as well as the piezo output (in blue) during the sequence:
\begin{itemize}
    \item{Stage 0:} The integral register of the piezo output is set to 1 V in order to reproducibly approach the cavity resonance from the high-voltage side.
    \item{Stage 1:} The controller attempts to stabilize the "reflection" signal to a detuning equal to 3 cavity bandwidths. As a consequence, the integrator of the piezo output immediately starts to drift towards lower voltages until the cavity resonance is nearly reached (at a voltage of 0.36 V). In order to avoid overshooting over the resonance, the gain factor is reduced to 0.001 during this stage. At the end of Stage 1, the cavity is thus safely stabilized close to resonance, in a region where the sign of the PDH error signal matches that of the laser detuning.
    \item{Stage 2:} After 630 ms in the previous state, the controller switches to a stabilization on the PDH error signal. The system thus quickly converges towards the stable attractor at 0-detuning. This is clearly visible on the reflection error signal which reaches its minimal level (see Fig. \ref{fig:lock_sequence}c).
\end{itemize}

The Lockbox even has an ``auto-lock'' option that monitors the error signals continuously, detects a potential loss-of-lock and automatically launches the re-locking sequence.

\subsection{Loop transfer-function measurement}
\label{sec:in_loop_transfer}

While a simple optimization based on trial and errors is often sufficient to tune the parameters of a PID controller, the knowledge of the total loop transfer function is of utmost importance in order to assess the true limitations and optimize the system. However, in many cases, some elements in the loop, such as piezoelectric actuators, have a convoluted frequency response that is hard to characterize separately. 
In this example, we use the network analyzer embedded in \pyrpl to probe the closed-loop transfer function of a laser-cavity lock, and deduce the complex frequency response of the piezoelectric actuator. 

\begin{figure}
    \centering
    \includegraphics[width=0.47\textwidth] {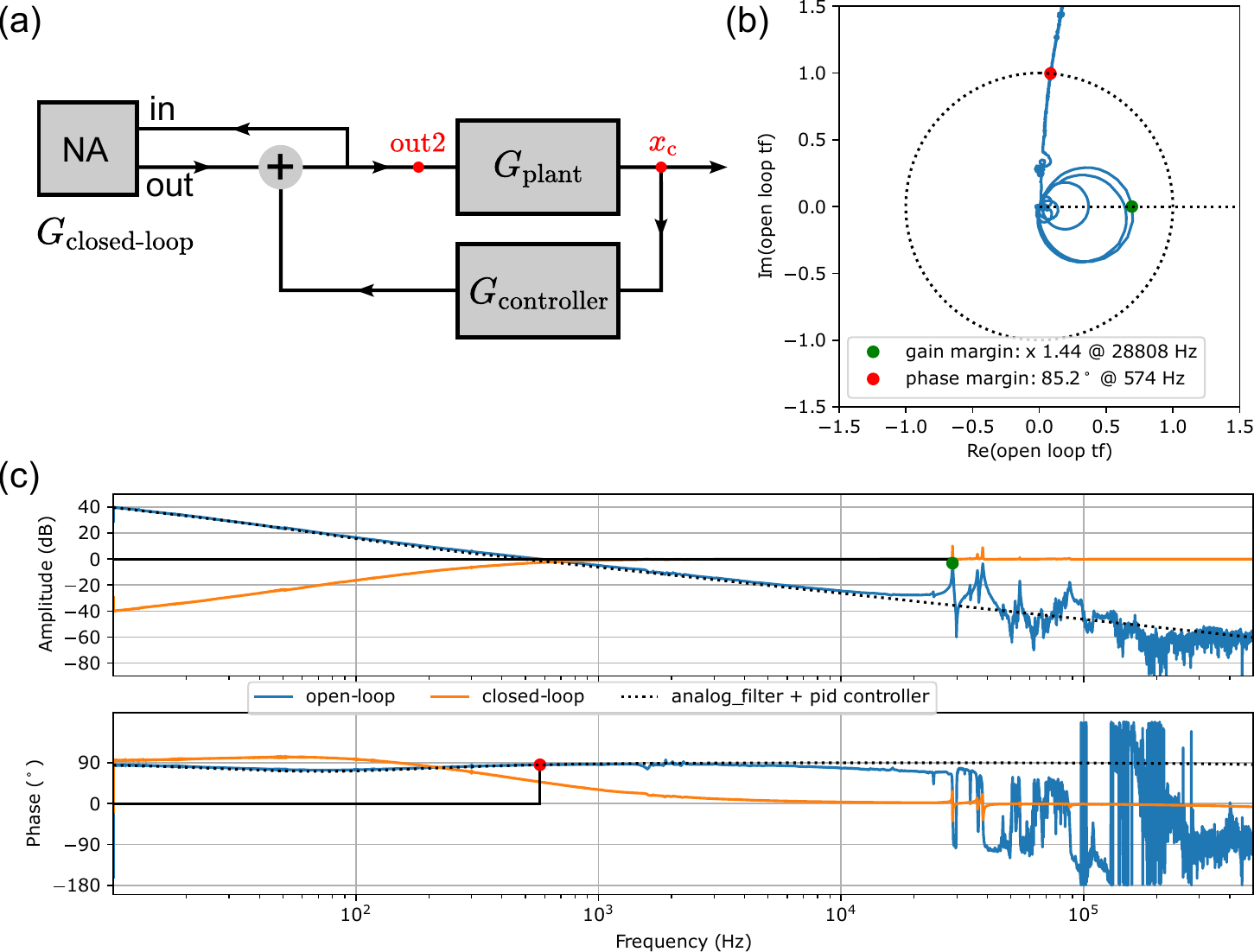}
    \caption{{\bf Measuring a loop transfer-function}.
    Generic feedback scheme (a) where the controller (here the RedPitaya) is acting on the plant (here an optical cavity). The Network Analyzer is adding a perturbation to the system and measures the total response including perturbation and counter-action from the controller. 
    This effectively measures the closed-loop transfer function $G_\mathrm{closed\_loop}$ (orange on (c)) of the system. The open-loop transfer function $G_\mathrm{open\_loop}$ (blue on (c)) is deduced using Eq.  \eqref{eq:closed_to_open_loop} of the main text. 
    The black dotted-line corresponds to the expected transfer-function of the loop, given by the product of the PID-controller and analog low pass filter transfer functions. 
    (b) The open-loop transfer function is represented on a Nyquist diagram where the gain margin and phase margin are represented by a green and red dot respectively. 
    The phase margin is limited by the phase-lag of the low-pass analog filter while the gain margin is limited by the piezo-electric actuator resonances. }
    \label{fig:transfer_function}
\end{figure}

To this end, we set the "output\_direct" of the network analyzer to the Red Pitaya output "out2" that is also used for feedback control (see the web page for full configuration instructions\cite{notebook}). Owing to the output summation stage of PyRPL, the signal fed to the piezo-electric actuator is thus the sum of the network analyzer modulation, and the feedback signal derived from the PDH error signal (see Fig. \ref{fig:transfer_function}a). The network analyzer input is set to "out2" as well, such that the measured transfer function $G_\mathrm{closed-loop}$ approaches 0 at low frequency, where the feedback signal nearly compensates the modulation, and 1 at high frequency where the gain of the controller is negligible. The closed-loop and open-loop transfer functions are linked via the relation:
\begin{equation}
    \label{eq:closed_to_open_loop}
    G_\mathrm{open-loop} = 1 - 1/G_\mathrm{closed-loop}.
\end{equation}
Fig. \ref{fig:transfer_function}c shows the measured transfer-function in orange and deduced open-loop transfer-function in blue. In order to accurately deduce the high-frequency open-loop behavior, where the $G_\mathrm{closed-loop}$ closely approaches 1, careful characterization of the background signal has to be performed (see the webpage  for details\cite{notebook}).

The open-loop transfer function can be decomposed as the product of three elements: the PID controller $G_\mathrm{controller}$, an analog low-pass filter $G_\mathrm{RC}$ with a cutoff frequency 50~Hz, and the electro-mechanical response of the piezoelectric actuator itself $G_\mathrm{piezo}$. 
The total open-loop transfer function matches very well the product $G_\mathrm{RC} G_\mathrm{controller}$ below 10 kHz. However, at higher frequency, mechanical resonances of the piezoelectric actuator manifest themselves as sharp peaks in the transfer function. 

The stability of the loop can be assessed in a straightforward manner from the Nyquist diagram represented in Fig. \ref{fig:transfer_function}b. In this plot, the total open-loop gain is represented as a parametric curve in the complex plane. The Nyquist stability criterion states that the loop is stable until the point (1, 0) doesn't get encircled by the response curve. We thus materialize the phase margin---the complex phase of the open-loop transfer-function at the unit-gain frequency---as a green dot. In this example, we have chosen to match the PI corner frequency of the PID controller with the analog low-pass filter cutoff, such that the phase margin approaches 90$^\circ$, as expected for a perfect integrator.
On the other hand, the gain margin---the maximum gain increase given the magnitude of the transfer-function at the 0-phase crossings---, materialized here by a red dot, is directly limited by the large piezo-electric resonances, which manifest themselves as large circles in this diagram. In the next section, we show how to use the IIR filter to compensate for the piezo-electric resonances and significantly increase the gain of the feedback loop without any instable behaviour.

\begin{figure*}[t]
    \centering
    \includegraphics[width=\textwidth] {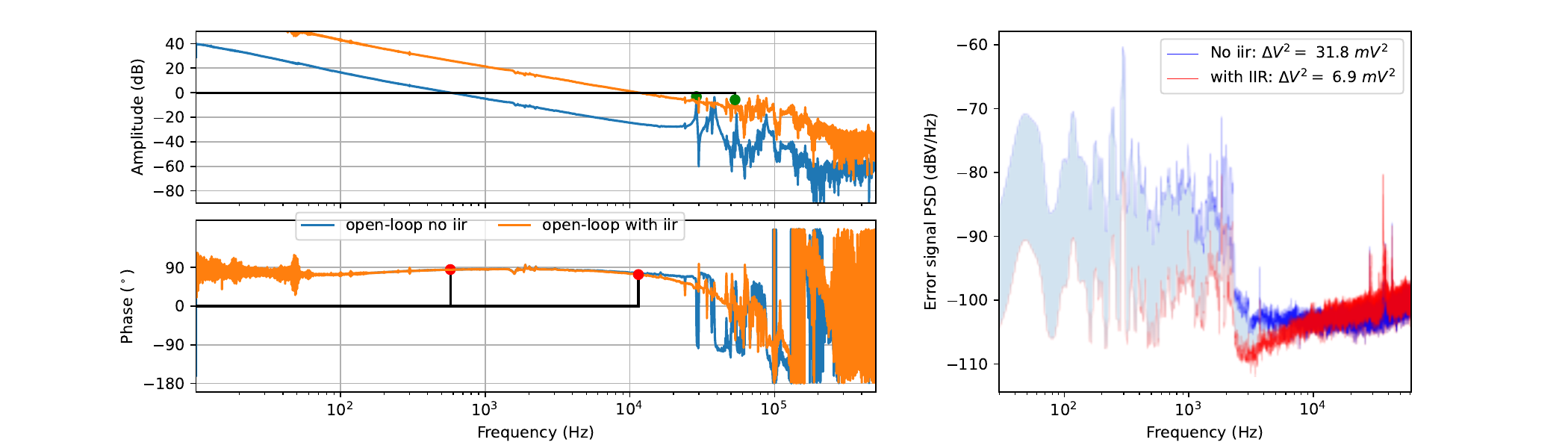}
    \caption{{\bf Optimized lock using the IIR filter}. (a) Open-loop transfer-function of the unoptimized lock described in Fig. \ref{fig:transfer_function} (blue curve) together with that of the optimized lock (orange) where the piezoelectric resonances have been numerically compensated with the IIR filter. The zeros and poles used for this particular filter implementation are listed in Fig. \ref{fig:iir_gui}. The $\sim23$~dB increase in overall loop gain corresponds to the unit-gain frequency increasing from $\sim500$~Hz to $\sim10$~kHz. 
    (b) Noise spectrum of the PDH error signal with (blue) and without (red) IIR filter. The variance (obtained by integrating the noise spectrum) decreases from $31.8$~mV$^2$ to $6.9$~mV$^2$ with the IIR filter. }
    \label{fig:iir_transfer_function}
\end{figure*}
\subsection{Lock optimization with IIR module}
\label{sec:iir_experiment}
The example of the previous section illustrates how the convoluted response of mechanical actuators tends to strongly impede the performances of control loops in optical systems. However, this limitation can be overcome: in fact, as long as the plant's response doesn't have any zero in the right-half plane---such a system is also called minimum-phase---it admits an invert that is causal and bounded, and that could thus be implemented in a specifically designed controller \cite{Bechhoefer2005}. 
However, the complexity of the required filter tends to divert the efforts of experimentalists towards mechanical engineering solutions, either attempting to push the mechanical modes towards higher frequencies or to damp the most problematic ones. Yet, digital control systems open an avenue for in-situ tuning of active compensation filters. Earlier work have demonstrated the compensation of the 10 main acoustic resonances of a piezoelectric actuator using a 25600 coefficients FIR filter \cite{Ryou2016}. 
Here, we demonstrate a similar result with the IIR module of PyRPL. 
As for the other examples discussed in this section, detailed instructions on how to tune the filter for the specific system to be stabilized are provided \cite{notebook}. We think that, together with the ability to characterize the actuator's transfer function in-situ (see section \ref{sec:in_loop_transfer}), this functionality should broadly popularize active cancellation techniques among the community.

Fig. \ref{fig:iir_gui} shows the GUI of the IIR module, once  fully configured to compensate the main resonances of the piezoelectric actuator characterized in section \ref{sec:in_loop_transfer}. To help with the filter design, the actuator's transfer function is displayed in the plot window (pink), together with the current filter's transfer function (in blue) and the product of these 2 curves (in cyan). The poles and zeros of the filter's transfer function are added and easily modified with a user-friendly graphical interface---zeros and poles are represented as dots and crosses respectively in the graph, and can be selected by a mouse click---until the amplitude response of the product curve is approximately flat. 
The final IIR filter is composed of 10 pairs of zeros and poles in the complex plane between 25 kHz and 90 kHz and is implemented with a delay of 10 cycles (see appendix \ref{AppendixIIR} for details). The $\sim 200~$ ns delay of the controller ($\sim100~$ns for the DAC and ADC, and $\sim 100~$ns due to the IIR filter pipeline itself) enable in practice to compensate mechanical resonances up to approximately 500~kHz. 
As visible in the second graph of Fig. \ref{fig:iir_gui}, the phase response of the product curve is also approximately flat as a function of frequency, indicating that the actuator is indeed close to a minimum phase system. 

We then proceed to close the loop with the designed filter and test the resulting performances. The total controller setup is composed of a PI controller and the IIR filter in series. In order to keep a similar phase margin as previously, we keep the same PI corner frequency, but increase the overall gain of the loop by a factor $\sim 14$. The transfer function of the loop is measured as previously and displayed in Fig. \ref{fig:iir_transfer_function}. As expected, the overall loop gain has been increased by 23 dB, while maintaining a gain margin of 2 (limited by a higher-frequency mode of the piezo-electric actuator at 60~kHz that is not compensated for). Accordingly, the unit-gain frequency is pushed by more than one decade from 500~Hz to 10~kHz thanks to the active compensation.

\subsection{Phase-locked loop}
Synchronizing two independent laser beams is a common need in various areas of experimental physics, including frequency comb stabilization, Doppler cancellation in optical fiber links, or optical time transfer. Generally, this is achieved by an optical phase-lock loop (PLL), where the difference frequency between a slave and a master laser is compared to a radio-frequency local oscillator. In practice, the local oscillator and the optical beatnote are sent into a phase detector which enables to stabilize the phase difference between these two signals. 

In the simplest form, a mixer can generate an error signal proportional to the sine of the phase difference. Such an "analog phase detector" can in principle detect arbitrarily small phase drifts. However, the monotonic range doesn't exceed $\pi$, which for one thing, limits the lock capture range (i.e. the maximum frequency difference to acquire lock), and for another, can lead to "phase-slips", whenever the lock fails to stabilize the phase below this critical value. 

A phase-frequency detector \cite{Prevedelli1995Phase}, consisting of a counter incremented upwards by the optical beatnote and downwards by the radiofrequency local oscillator has a much larger monotonic range (determined by the saturation value of the counter), but this comes at the expense of a limited precision as it is only sensitive to the edges of the input signals, rather than their phase integrated over a full period \cite{beverini2004analog} (the situation is even worse in a digital signal processing environment where edge detection is time-binned by an external clock). 

All digital PLLs \cite{Kumm2010}, on the other hand, proceed by explicitly evaluating the phase difference between the 2 input signals. Since this signal is inherently defined on a $2\pi$ interval, it needs to be phase-unwrapped in order to extend the range of the phase signal. As a consequence, such phase detectors feature a precise and linear behavior over a large range. 

In this example, we use the CORDIC phase estimator of \pyrpl to phase-lock 2 Mephisto YAG lasers. The algorithm is based on a binary estimation of the phase knowing the I and Q coordinates of the demodulated signal.  First, the signs of I and Q give the quadrant the phase lies in. The implementation in PyRPL includes an overflow procedure when the phase jumps from the last quadrant to the first (resp. the first to the last) and increment (resp. decrement) a turn counter.  The IQ vector is then rotated in the $-\frac{\pi}{4}$, $\frac{\pi}{4}$ range and the test condition of the binary search is the sign of the vertical coordinate Q of the vector. The algorithm recursively applies pseudo rotation matrices with decreasing magnitudes, in the clockwise or counter-clockwise direction depending on the sign of Q and accumulate the corresponding angle. The matrix coefficients are of the form $2^{-n}$ allowing a single cycle computation  with additions and bit-shifts. The phase is coded on a 14-bit register, the 4 most significant bits are the turn count which is 2 bits wide allowing to encode phase between $-4\pi$ and $4\pi$, and 2 more bits for the quadrant. Follows 9 stages of recursive search to complete the 10 least significant bits with a quantization error of 0.11°. With 14 bits in a $8\pi$ range, the theoretical quantization is $8\pi / 2^{15}=0.044$°. This could be approached by a tenth stage in the algorithm but would require extra bits for all registers and a careful truncation of the result. The precision is also spoiled by a factor $4/\pi$ due to the use of the non optimal rotation angles $\tan^{-1} (2^{-k})$ instead of $(2^{-k})\frac{\pi}{4}$.

\begin{figure*}
\centering
    \includegraphics[width=0.9\textwidth] {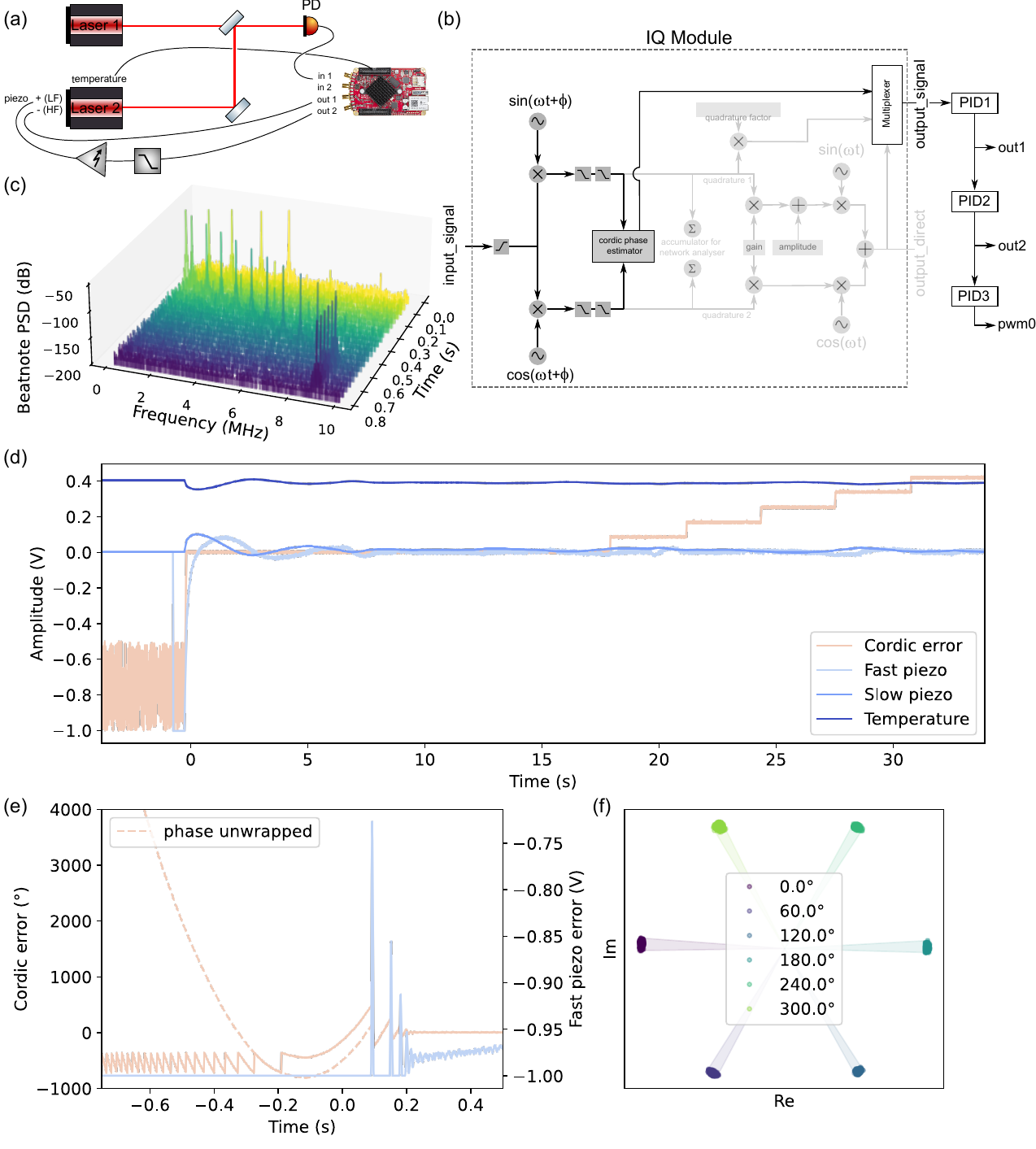}
    \caption{{\bf Implementation of a Phase-Lock Loop}. (a) Schematics of the optical and analog electronic setup: out1 and out2 are connected to fast and slow inputs of the piezoelectric laser actuator respectively. The pulse-width modulation output PWM0 is connected to the crystal temperature controller. The beatnote signal obtained by superimposing the two laser beams is acquired on a photodiode connected to in1. (b) Schematic of the IQ module used to compute the phase difference between the two lasers. The signal is demodulated at the desired beatnote frequency, and the angle in the rotating frame is estimated from the two quadratures via the CORDIC phase estimator. The error signal is then sent to cascaded PID filters, that control the laser actuators: the output of PID1, used for the fast piezoelectric actuator is maintained in the middle of its range by being fed to the slow piezoelectric controller (PID2). In turn, the output of PID2 is fed to the input of the laser temperature controller (PID3) for long term stabilization. (c) Spectrogram of the beatnote signal during lock acquisition: as the lock is switched on, the beatnote frequency quickly drifts towards the desired value of 9 MHz. (d) Error signals during the lock acquisition sequence: the lock is enabled at $t=0$, after a short transient, the CORDIC error signal, as well as all outputs except for the laser temperature stabilize around 0 V. At $t=18$~s, the phase setpoint is varied stepwise by increments of 60$^\circ$ to demonstrate the ability of the setup to stabilize the beatnote around an arbitrary phase. (e) Zoom on the lock-acquisition sequence around $t=0$. Curve colors are identical to those of panel (d). In addition, we estimate the unsaturated evolution of the beatnote phase by unwrapping the CORDIC error signal (dashed curve). (f) IQ diagram of the demodulated beatnote amplitude for the various setpoints. The standard deviation of the phase in each step is approximately 0.5$^\circ$. }
    \label{fig:pll}
\end{figure*}
The beatnote between the 2 lasers is directly fed into an analog input of the Red Pitaya and demodulated at 9 MHz by an IQ module. The output multiplexer of the IQ module is set to the CORDIC phase estimator and fed to the actuators of the slave laser via several PID modules (see Fig. \ref{fig:pll} and the notebooks\cite{notebook} for configuration details). 
In order to simultaneously achieve a large actuation bandwidth and dynamic range while minimizing the electronic and digitization noise, we use three distinct actuators: the output "fast piezo" is directly connected to one of the terminals of the piezoelectric actuator, yielding a bandwidth of 100 kHz, the output "slow piezo" is connected to the other piezo terminal via a high-voltage amplifier and analog low-pass filter (10 Hz). Finally, the output "temperature" controls the temperature of the YAG crystal via a pulse-width modulation output of the Red Pitaya (see Fig. \ref{fig:pll}a). The PID modules are connected in series, such that the output of the fastest actuators are maintained close to the middle of their range by the action of the slower actuators. 

Fig. \ref{fig:pll}d shows the time evolution of the various error signals during the lock acquisition sequence. During the first part of the sequence, for times $-1$~s~$\le t \le$~0~s, only the fast piezo-lock is enabled. The beatnote frequency is still far from that of the local oscillator, such that the demodulated signal rotates quickly in the IQ plane. 
Even though the CORDIC error signal saturates at $\pm 4 \pi$, the signal is reset to $\pm 2 \pi$ upon overflow. As a consequence, the error signal has a correct sign durign this stage, with an average value of $\pm 3 \pi$, and the fast-piezo signal is saturated at -1 V. 
At time $t=0$, the other lock-channels are activated simultaneously to acquire the lock. 
Fig. \ref{fig:pll}c shows the evolution of the beatnote spectrum during this stage. As the frequency $\nu_2$ of laser 2 continuously drifts towards $\nu_1 + 9$ MHz, the frequency of the beatnote $|\nu_2 - \nu_1|$ undergoes a sharp direction change when $\nu_2 = \nu_1$. After 0.6 s, the beatnote is stabilized to the desired value, with all error signals centered in the middle of their range, except for the laser temperature. 
Fig. \ref{fig:pll}e shows a zoom around the lock-acquisition moment. 
Eventhough the CORDIC error signal has a discontinuous behavior upon saturation, it can be unwrapped in post-processing to yield the continuous phase-evolution of the beatnote (dashed line). 

At time $t=18$~s, the phase setpoint of the lock is incremented by steps of $60^\circ$ to demonstrate the ability of the loop to stabilize the beatnote at arbitrary phases. 
Fig.~\ref{fig:pll}f shows the evolution of the two quadratures as a parametric plot in the compex-plane. 
The standard deviation of the individual blobs indicates a precision of $0.5^\circ$ for the stabilization of the beatnote.

\section{Conclusion}

In conclusion, the open-source software package PyRPL, is a streamlined and pragmatic tool for implementing automatic digital feedback controllers in quantum optics experiments. The transition from traditional analog controllers to this digital approach presents several advantages.
Significant cost reductions are achieved as typically a single Red Pitaya board can replace an array of equipment including signal generators, sweep generators, PID controllers, and analog mixers. Importantly, digital operation does not result in any detectable performance loss due to digitization noise. Furthermore, the approach effectively bypasses the issue of fluctuating offsets that can plague systems reliant on analog demodulation.
A standout feature of PyRPL is its modular architecture which, when paired with a high-level programming interface and GUI, enables the agile development and debugging of complex feedback systems. 
Once a feedback loop has been configured, it is typically operated in a push-button manner owing to the lock-acquisition logic included in PyRPL. 
Finally, PyRPL's open-source nature allows it to be easily tailored to specific needs and enables constant refinement by a growing user community. This adaptability and continual evolution ensure PyRPL remains a dynamic and valuable tool in the field of quantum optics experiments.

\begin{acknowledgments}
The authors thank Pierre Clad\'e for extended discussions. They acknowledge support from ANR projects ExSqueez (ANR-15-CE30-0014) and QFilters (ANR-18-JSTQ-0002). L.~N. acknowledges support from the FP7-cQOM Initial Training Network. S.~C. acknowledges support from a Marie Slodowska-Curie post-doctoral fellowship (project 660941 - SQZOMS). M. C. and P.-E. J. acknowledge support from a PhD grant of 'R\'egion \^Ile-de-France' (SIRTEQ projects OSLO and QuBeat).
\end{acknowledgments}

\section*{Data Availability Statement}

The data that support the findings of this study are openly available in {\url{https://github.com/lneuhaus/pyrpl/tree/main/docs/example-notebooks/article\_examples}, reference number \cite{notebook}.

\appendix

\section{Infinite Impulse Response filter\label{AppendixIIR}}

 The following theoretical discussion closely follows Ref. \cite{Oppenheim1975}.
Causality ensures the output of a filter at cycle $n=n_0$ may only depend on the input signal for $n<n_0$: 
 \begin{equation}
y(n) = \sum_{i=0}^\infty x(n - i)h(i).\end{equation} 
 A useful representation of the impulse response $h$ is given by its Z-transform:
\begin{equation}
H(z) = \sum_{n=0}^{\infty} h(n) z^{-n},
\label{z-transform}
\end{equation} 
which is defined within a region of convergence $|z|>r$. Restricting our attention to stable filters, where bounded input signals lead to bounded outputs, we have $r<1$ and the system can be fully described equivalently by its frequency response:
\begin{equation}
H(e^{j \omega_r}) = \sum_{n=0}^{\infty} h(n) e^{-j \omega_r n}.
\label{fourier-transform}
\end{equation} 
In the above formula, the radian frequency $\omega_r$ takes values in the interval $[-\pi, \pi]$, and it is linked to the usual continuous-time angular frequency $\omega$ by the relation $\omega_r =\omega T$\footnote{It follows from this relation that the transfer function of a discrete-time filter is periodic, with a period of twice the 
Nyquist frequency $1/T$.}. 

Most practical filters can be represented by a rational Z-transform over the region of convergence:
\begin{equation}
H(z) =\frac{P(z)}{Q(z)},
\label{polynom}
\end{equation} 
where $P(z)$ and $Q(z)$ are polynomials in $z$. We denote $z_i$ and $p_i$ the zeros of the numerator and denominator respectively, hence obtaining:
\begin{equation}
H(z) = k \frac{\prod_{i=1}^{M} (1 - z_i z^{-1})}{\prod_{j=1}^{N}{(1 - p_i z^{-1}})}.
\label{factored}
\end{equation}
For feedback applications, we will restrict our discussion to proper IIR filters that obey $M\le N$, since improper filters with $M > N$ are expected to result in an unstable closed-loop behavior at high-frequencies where unavoidable delays occur (e.g. from the ADCs and DACs), thus impacting the phase margin. Moreover, to ensure a real time-domain representation $h(n) \in \mathds{R}$, any non-real pole or zero must be accompanied by its complex conjugate in Eq. (\ref{factored}). 

In summary, a practical IIR filter can be defined unambiguously by the lists of zeros and poles $\{z_i\}_{1\le i \le M}$ and 
$\{p_j\}_{1\le j \le N}$ with $M\le N$ and the overall gain factor $k$. In the next section, we describe how one can implement an arbitrary filter, with a transfer function described by Eq. \eqref{factored} in a real-time signal processing environment.

\subsubsection*{IIR filter implementation}

When designing an IIR filter, we specify the desired transfer function as a list of poles and zeros, together with a prefactor $C$ corresponding to the gain at zero-frequency. In order to facilitate the interpretation in terms of actual frequency, the user specifies the zeros and poles $\{\tilde z_i\}_{1\le i \le M}$ and $\{\tilde p_j\}_{1\le j \le N}$ in terms of continuous time Laplace frequencies  $s = i \omega + \gamma$\footnote{A common source of confusion comes from the fact that the Laplace transform maps the physical frequencies onto the imaginary axis for the parameter $s$, hence, the characteristic frequencies $\omega$ of the transfer function features are described 
by the imaginary part of the poles/zeros, while their typical
linewidth $\gamma$ is given by their real part.}.  
Starting from these inputs, the arithmetic manipulations necessary for the practical implementation of the filter on the FPGA board are described in a public python file \footnote{\url{https://github.com/lneuhaus/pyrpl/blob/master/pyrpl/
hardware_modules/iir/iir_theory.py}}. The main steps are summarized here:
\begin{enumerate}
    \item The function $\mathrm{IirFilter.proper\_sys}$ ensures that all non-real zeros/poles have a conjugate partner. Otherwise, the missing items are appended to the list. Once this operation has been completed, if the number $M$ of zeros  exceeds the number  $N$ of poles, extra-poles with a large real Laplace frequency are appended to the list to make the system strictly proper. 
    Contrary to most other DSP operations in PyRPL, the calculation of the IIR filter output $y(n)$ is too complex to be carried out in a single FPGA cycle. Instead, a number of cycles $N_\mathrm{loops} = \left \lceil{N/2} \right \rceil$ is required. Since the effective sampling time is given by $T_\mathrm{IIR} =  N_\mathrm{loops} T$, it is important to determine $N_\mathrm{loops}$ at such an early stage in the process.

    \item The function $\mathrm{IirFilter.rescaled\_sys}$ rescales the zeros and poles in terms of angular frequency, and calculates the prefactor $k$ of Eq. \eqref{factored} as a function of the specified DC-gain $C$.
    
    \item Before implementing the filter, we convert the continuous time zeros and poles into discrete-time zeros and poles of the Z-transform via the mapping:
    \begin{align*}
    z_i &= e^{\tilde z_i T_\mathrm{IIR}} \\
    p_i &= e^{\tilde p_i T_\mathrm{IIR}}.
    \end{align*}  
    
    \item The function $\mathrm{IirFilter.residues}$ performs the partial fraction expansion of the desired Laplace transform using the Heaviside cover-up method:
\begin{align}
\label{factored_laplace}
H(z) &= k \frac{\prod_{i=1}^{M} (1 - z_i z^{-1})}{\prod_{j=1}^{N}(1 - p_j z^{-1})} \\
H(z)&= D +  \sum_{j=1}^{N}\frac{ r_j}{1 - p_j z^{-1}} \,,
\label{residues_laplace}
\end{align}
    where $D$ is a constant for proper filters ($M\le N$)  and is non-zero only for strictly proper filters ($M=N$). The residues $\{r_j\}_{1 \le j \le N}$, zeros $\{z_j\}_{1 \le j \le N}$, and constant feed-through $D$ are an unambiguous representation of the filter. Furthermore, contrary to the factored representation \eqref{factored_laplace} involving the poles, zeros, and gain $k$, 
    an expanded form such as Eq. \eqref{residues_laplace} lends itself naturally to a modular implementation where the filter output is obtained by summing the outputs of different modules, provided each term in the Z-transform can be implemented by relatively simple DSP operations.

\item In order to implement a discrete time filter with the Z-transform described by Eq. \eqref{residues_laplace}, we combine pairs of terms after the summation symbol into second-order sections:
\begin{align}
H(z) &= D +\sum_{j=1}^{N'} \frac{b_0 + b_1 z^{-1}}{1 + a_1 z^{-1} + a_2 z^{-2}},\\
\mathrm{with\,\,\,} b_0 &= r_j + r_j', \,\,\, b_1 = -p_jr_{j'}-p_j'r_j,\\
a_1 &= -p_j - p_j', \,\,\, 
a_2 = p_j p_j'.
\label{biquads_decomposition}
\end{align}
By making sure that complex poles are paired with their complex conjugate ($p_j'=\bar p_j$), we ensure that all coefficients $b_0, b_1, a_1, a_2$ are real as residues $r_j$ and $r_j'$ are also complex conjugate. Each term in the sum \eqref{biquads_decomposition} is a simple IIR filter that can be physically implemented by calculating the output at cycle $n$ from the current and previous inputs $x(n)$ and $x(n-1)$ and previous output values, $y_j(n-1)$ and $y_j(n-2)$:
\begin{align*}
y_j(n) =& b_0 x(n) + b_1 x(n-1) \\
&- a_1 y_j(n-1) - a_2 y_j(n-2).
\end{align*}
Indeed, by moving the 2 last terms to the left-hand side and by taking the Z-transform of the equation, we get:
\begin{equation}
Y_j(z) [1 + a_1 z^{-1} + a_2 z^{-2}] = X(z) [b_0 + b_1 z^{-1}].
\end{equation}
Hence, the Z-transform of the second-order section $y_j$ is given by:
\begin{equation}
H_j(z) = \frac{Y_j(z)}{X(z)} =  \frac{b_0 + b_1 z^{-1}}{1 + a_1 z^{-1} + a_2 z^{-2}}.
\end{equation}
The constant feedthrough term D is implemented as a separate second-order section with $b_0=D$, and $a_1=a_2=b_1=0$. 

All the coefficients $b_0$, $b_1$, $a_1$ and $a_2$ for the individual biquads filters are calculated by the function $\mathrm{IirFilter.rp2coefficients}$.

\item{Since the second-order sections are sequentially implemented in the FPGA, we rearrange the different biquads by ascending frequency in order to minimize the delay experienced by the sections with the largest frequencies in the function $\mathrm{IirFilter.minimize\_delay}$.}

\item{In the function $\mathrm{IirFilter.coefficients\_rounded}$, we convert the floating number coefficients calculated above into fixed-point precision coefficients, and check for the magnitude of rounding errors. An error message is prompted when the relative error is too large.}
\end{enumerate}

In the Verilog code, we only implement a single biquad filter module, which computes the output for the coefficients of each second-order section in subsequent FPGA clock cycles. The cumulative sum of all second-order section outputs is the filter output. 
As mentioned above, this implementation leads to a reduced effective sampling rate equal to the FPGA clock rate divided by the number of second-order sections, i.e. by half the filter order. Filter coefficients are represented as fixed-point numbers with 3 (resp. 29) bits before (resp. after) the radix point. As a compromise between filter complexity and FPGA resources devoted to the IIR filter, we allow for a maximum filter order of 28, i.e. 14 second-order sections. The filter is preceded by a first-order low-pass filter to avoid aliasing when the sampling rate is significantly lower than the FPGA clock frequency. 

\medskip
\section*{References}

%

\end{document}